\documentclass[a4paper]{cas-dc}
\let\printorcid\relax % Remove ORCID footnote

\usepackage{graphicx}
\usepackage{amsmath}
\usepackage{amssymb}
\usepackage{setspace}
\usepackage{epstopdf}
\usepackage{color}
\usepackage[ruled, linesnumbered, vlined]{algorithm2e}
\usepackage{algpseudocode}
\usepackage{enumitem}
\usepackage{multirow}
\usepackage{cite}
\usepackage{siunitx,etoolbox} % table font bond
\usepackage{subfigure}
\usepackage{subcaption}
\usepackage{physics}
\usepackage{soul}
\newtheorem{rmk}{Remark}
\usepackage{array}
\usepackage{booktabs}
\usepackage{tabularx}  % works for two columns
\usepackage{makecell}
\usepackage{soul}
\usepackage{caption}

\def\tsc#1{\csdef{#1}{\textsc{\lowercase{#1}}\xspace}}
\tsc{WGM}
\tsc{QE}
\begin{document}
\let\WriteBookmarks\relax
\def\floatpagepagefraction{1}
\def\textpagefraction{.001}

\shorttitle{Hybrid model-based EMPC for shipboard carbon capture process}    

\shortauthors{Xuewen. Zhang et al.}  

\title [mode = title]{Machine learning-based hybrid dynamic modeling and economic predictive control of carbon capture process for ship decarbonization}  




\affiliation[1]{organization={School of Chemistry, Chemical Engineering and Biotechnology, Nanyang Technological University},
            addressline={62 Nanyang Drive}, 
            % city={},
%          citysep={}, % Uncomment if no comma needed between city and postcode
            postcode={637459}, 
            country={Singapore}}

\affiliation[2]{organization={Nanyang Environment and Water Research Institute (NEWRI), Nanyang Technological University},
            addressline={1 CleanTech Loop}, 
            % city={},
%          citysep={}, % Uncomment if no comma needed between city and postcode
            postcode={637141}, 
            country={Singapore}}

\affiliation[3]{organization={Systems Engineering, Cornell University},
            addressline={Ithaca}, 
            postcode={14853}, 
            city={New York},
%          citysep={}, % Uncomment if no comma needed between city and postcode
            state={NY},
            country={USA}}

\author[1]{Xuewen Zhang}

\author[1]{Kuniadi Wandy Huang}

\author[1]{Dat-Nguyen Vo}

\author[1,2]{Minghao Han}

\author[3]{Benjamin Decardi-Nelson}

\author[1,2]{Xunyuan Yin}\cormark[1]

\cortext[1]{Corresponding author: X. Yin. Tel: (+65) 6316 8746. Email: xunyuan.yin@ntu.edu.sg.}


% \nonumnote{}

\begin{abstract}
    Implementing carbon capture technology on-board ships holds promise as a solution to facilitate the reduction of carbon intensity in international shipping, as mandated by the International Maritime Organization. In this work, we address the energy-efficient operation of shipboard carbon capture processes by proposing a hybrid modeling-based economic predictive control scheme. Specifically, we consider a comprehensive shipboard carbon capture process that encompasses the ship engine system and the shipboard post-combustion carbon capture plant. To accurately and robustly characterize the dynamic behaviors of this shipboard plant, we develop a hybrid dynamic process model that integrates available imperfect physical knowledge with neural networks trained using process operation data. An economic model predictive control approach is proposed based on the hybrid model to ensure carbon capture efficiency while minimizing energy consumption required for the carbon capture process operation. The cross-entropy method is employed to efficiently solve the complex non-convex optimization problem associated with the proposed hybrid model-based economic model predictive control method. Extensive simulations, analyses, and comparisons are conducted to verify the effectiveness and illustrate the superiority of the proposed framework.
\end{abstract}


\begin{highlights}
    \item \textcolor{black}{A ship decarbonization design is formulated to reduce CO$_2$ emissions from ship fuel combustion engines by integrating the post-combustion carbon capture process and seawater cooling with the ship energy generation system.}\vspace{-5pt}

    \item \textcolor{black}{A machine learning-based hybrid modeling method is proposed for the shipboard carbon capture process, with simulation results showcasing good performance in capturing the complex dynamics of the shipboard carbon capture process under varying engine load demands.}\vspace{-5pt}

    \item \textcolor{black}{An economic model predictive control design is proposed based on the developed hybrid model considering the operating cost of the shipboard carbon capture process, with the cross-entropy method being employed to efficiently solve the complex optimization problem involved.}\vspace{-5pt}

    \item \textcolor{black}{The integration of the developed learning-enabled hybrid model and the proposed economic model predictive control streamlines carbon capture process operations, leading to energy-efficient operations with a small impact on carbon capture rate.}
\end{highlights}

%\nocite{*}

% Keywords

\maketitle

\section{Introduction}

The transportation sector is one of the most substantial contributors to climate change, with international shipping alone responsible for around 3\% of the world's greenhouse gas (GHG) emissions as of 2022~\cite{bach2023imo}. Common marine fuels, such as heavy fuel oil, diesel, and liquefied natural gas, continue to be widely utilized, and the combustion of these fuels produces substantial carbon emissions~\cite{bicer2018clean}. Projections indicate that CO$_2$ emissions from shipping transport will increase to 1.6 billion tons by 2050~\cite{capros2013eu}. The International Maritime Organization (IMO) has implemented mandatory measures to reduce GHG emissions within international shipping and has revised its strategy for GHG emission reduction in 2023~\cite{mepc20232023}. The revised strategies include an ambition to reach net-zero GHG emissions from international shipping by 2050, and outline an indicative checkpoint to reduce the total annual GHG emissions from international shipping by at least 70\% by 2040~\cite{mepc20232023}.

One approach to reduce CO$_2$ emissions from ships is transitioning to clean fuels~\cite{sofiev2018cleaner, bicer2018clean}. Particularly, carbon-free fuels like hydrogen and ammonia are suitable to replace traditional marine fuels~\cite{bicer2018clean}. The adoption of clean fuels faces challenges in real-world applications, notably in the storage and transportation of compressed or liquefied forms~\cite{aakko2023reduction}. Moreover, combustion technology is still in its premature stages, and modifying/upgrading power engine systems is anticipated to necessitate substantial investment in both economic and infrastructural aspects. Therefore, from a cost competitiveness perspective, significant challenges persist, and the transition to clean fuels in international shipping is unlikely to be completed in the near future~\cite{wang2022review}.

Another promising strategy is to use post-combustion carbon capture (PCC) system on-board ships to reduce ship carbon emissions due to its ability to be easily retrofitted into existing systems without significant changes~\cite{aakko2023reduction}. Various studies have investigated post-combustion carbon capture for ships~\cite{feenstra2019ship, luo2017study,ros2022advancements, bayramouglu2023application, vo2024advanced}. \textcolor{black}{In~\cite{feenstra2019ship}, a ship-based carbon capture plant for diesel or LNG-fuelled vessels using MEA or piperazine solvents was analyzed. In~\cite{luo2017study}, a PCC plant was developed for a typical cargo ship equipped with two four-stroke reciprocating engines. With a waste heat recovery (WHR) system, a 73\% carbon capture rate was achieved at 85\% load on the diesel engines. In~\cite{ros2022advancements}, based on the results of the DerisCO2 project, advancements in ship-based carbon capture technology were discussed for LNG-fuelled ships, focusing on solvent selection, heat integration, and the effect of ship movement. In~\cite{bayramouglu2023application}, an MEA solvent-based carbon capture process with a WHR system was developed; this work has highlighted the dependence of the carbon capture rate on varying engine loads and various parameters of the carbon capture plant, including MEA concentration and the inlet temperature of the absorption column. In~\cite{vo2024advanced}, four advanced designs for shipboard carbon capture process were proposed and analyzed for energy-efficient operation using correlation analysis and machine learning-based optimization. In the shipboard PCC plant, ship engines combust fuel to generate propulsion power, producing flue gases in the processes.} These flue gases are directed to a shipboard PCC plant for treatment before they are discharged into the atmosphere. Within the PCC process, regenerating captured CO$_2$ demands a substantial amount of heat energy~\cite{decardi2018improving}. In this research, we consider onboard PCC for ship decarbonization, and propose to use a separate diesel gas turbine and a WHR system to provide heat energy for regenerating CO$_2$ from the rich solvent.

Variations in the ship engine load lead to significant fluctuations in the flue gas flow rate, which also affect the recovery of heat energy from the WHR system and the load of the diesel gas turbine. The aforementioned intricacies, coupled with the complex structure and large scale of the integrated shipboard PCC plant, present challenges in attaining safe and energy-efficient process operation while maintaining a high level of carbon capture rate. These considerations highlight the importance of developing and implementing advanced control solutions for consistent and efficient operation of shipboard carbon capture processes.

Model predictive control (MPC) is one of the most widely used advanced control methodologies~\cite{morari1999model}. The development of a successful MPC scheme requires a high-fidelity model that can accurately characterize the dynamic behavior of the underlying system/industrial process. In the existing literature, MPC designs have been proposed for land-based PCC processes. In~\cite{sahraei2014controllability, he2016flexible}, simultaneous scheduling and MPC frameworks were developed for a PCC plant, where linear MPC designs were proposed based on identified models. In~\cite{jung2021model}, a set-point tracking MPC approach was proposed for a PCC plant with an advanced flash stripper and compared with decentralized proportional-integral-derivative controllers. In~\cite{zhang2016development}, a linear MPC scheme was implemented for a PCC plant attached to a supercritical coal plant, and the performance of this control design was evaluated and compared with a proportional-integral-derivative controller under different possible conditions.  In~\cite{zhang2018nonlinear}, a nonlinear MPC method was proposed based on a nonlinear additive autoregressive with exogenous variables model, which demonstrated superior performance in comparison with the linear MPC design proposed in~\cite{zhang2016development}. In~\cite{akinola2020nonlinear}, a nonlinear autoregressive with exogenous input (NARX) model was established using the forward regression orthogonal least square - error reduction ratio algorithm to improve the model accuracy, and a nonlinear MPC design was proposed based on the NARX model; this method provided improved performance compared with a linear MPC. In~\cite{panahi2012economically}, a robust MPC was proposed for a land-based PCC plant based on an identified dynamic model, and was compared with decentralized proportional-integral controllers. In~\cite{patron2022integrated}, a nonlinear MPC was proposed based on a first-principles model of a PCC plant, and an operational scheme that includes the nonlinear MPC, moving horizon estimation, and real-time optimization was developed. In~\cite{decardi2018improving}, a first-principles model for the PCC plant was developed, and economic MPC was proposed; this method demonstrated the capability to reduce the overall operational cost as compared to set-point tracking MPC.

\textcolor{black}{However, it may not be appropriate to directly apply the existing control methods developed for land-based PCC plants to the shipboard applications. The transfer of existing control designs from land-based to shipboard applications faces two key challenges: 1) significant differences in the dynamic behaviors of the two types of plants; 2) the differences in the objectives in real-time process operations. Specifically, land-based PCC plants that connected with a power plant are different from shipboard PCC plants in terms of process designs, component sizing, and manner of heat supply~\cite{luo2017study}. Due to spatial constraints on ships, the components of the PCC plant, such as the absorption and desorption columns, are designed to be smaller than those used in land-based PCC systems~\cite{luo2017study}. The heat supply for the shipboard PCC plant is sourced from the ship engines instead of a power plant for land-based applications. Additional heat sources are typically required for the operation of shipboard PCC plants~\cite{luo2017study, ros2022advancements, bayramouglu2023application }. In addition, different ship operational conditions lead to variations in the ship engine load, which significantly impact the heat energy supplied to the PCC plant and further affect the dynamic behavior of the shipboard PCC plant~\cite{luo2017study, ros2022advancements}. The ship engine load is considered a known disturbance to the shipboard PCC plant, which is another unique feature of the shipboard PCC plant as compared to the land-based PCC plant. Furthermore, the shipboard PCC plant considered in this work utilizes seawater for cooling during voyages, and the seawater flow rate is one of the control inputs. Ships experience constantly changing operational and environmental conditions, suh as the ship engine load and seawater state. Therefore, these above-mentioned differences and features collectively lead to dynamic behaviors that are significantly different from those of land-based counterparts, which typically operate in stable environments with steady inputs and outputs. This poses challenges in directly applying the existing control algorithms developed for land-based PCC to shipboard PCC plants. Additionally, we note that reducing carbon emissions and achieving energy-efficient operations are both important for PCC plants on-board ships, especially under varying operational conditions. Therefore, an appropriate control objective for shipboard PCC plants should prioritize optimizing the energy efficiency under varying engine loads and minimizing the impact of energy optimization on the carbon capture rate. Traditional control strategies for land-based PCC plants were primarily designed to address set-point tracking tasks~\cite{zhang2018nonlinear, akinola2020nonlinear, zhang2016development}; therefore, they cannot be directly applied to improve the energy efficiency of the operation of shipboard PCC plants, and may not be best suited for real-time control of these PCC plants on-board ships. In~\cite{han2025deepneuralkoopman}, a purely data-driven advanced control method was proposed for shipboard PCC. Specifically, a data-driven linear adaptive Koopman model was built to characterize the economic costs associated with real-time process operation and to predict several key output variables. Then, a convex optimization-based control scheme was developed for efficient process operations~\cite{han2025deepneuralkoopman}. As compared to the current work, the method in~\cite{han2025deepneuralkoopman} is purely data-driven, and it does not provide predictions for the full-state of the process; instead, it considers only the output variables that are directly associated with the economic cost and the output variables that are subject to hard constraints. Additionally, the valuable available physical information was not utilized in~\cite{han2025deepneuralkoopman}.}

% The significant disparities from the land-based counterpart in terms of process designs, component sizes, manner of heat supply, among other factors~\cite{luo2017study}, pose challenges in transferring these existing control methodologies from land-based to shipboard applications. Additionally, there have been no results on dynamic modeling and control of shipboard post-combustion carbon capture plants.

Based on the above observations, in this work, we aim to address the dynamic modeling and optimal control problem for the shipboard post-combustion carbon capture process. While the existing results for land-based facilities may offer insights into the dynamic modeling of shipboard carbon capture plants, we recognize the limitations of these methods when considering the application to shipboard plants: 1) first-principles modeling as adopted in~\cite{decardi2018improving, harun2012dynamic,yin2020IJACSP} can encounter challenges due to the unavailability of the values of some crucial model parameters; 2) system identification, as illustrated in~\cite{sahraei2014controllability, he2016flexible, zhang2016development, zhang2018nonlinear, akinola2020nonlinear, panahi2012economically}, often relies solely on input-output data, leading to models that overlook the dynamic behaviors of key state variables. Considering the importance of ensuring safety and energy efficiency in shipboard applications, constraints on key state variables should be enforced, and dynamic information on extensive states is needed to calculate and reduce operational costs. System identification-based approaches typically do not have sufficient capability to make well-informed control decisions for shipboard applications, considering both safety and economic efficiency in process operation. To handle the large scale and complex structure and dynamics of the integrated plant, a promising modeling framework is hybrid modeling, which seamlessly integrates the available first-principles knowledge and data information. Hybrid models have the potential to offer high prediction accuracy, strong generalization capabilities, and good interpretability~\cite{shah2022deep,bangi2020deep}. In~\cite{ghosh2019hybrid}, a data-driven model, using the batch subspace identification method, was developed for the residuals between the outputs generated by the plant and first-principles model.  In~\cite{oliveira2004combining, su1993integrating}, machine learning was employed to complement inaccurate first-principles models. In~\cite{shah2022deep, bangi2020deep, hosen2011control, psichogios1992hybrid}, machine learning was employed to estimate some parameters of the first-principles models. Based on a dynamic model, advanced control can be developed within the framework of MPC to optimize the real-time operation while ensuring various physical constraints are satisfied. Considering the significant fluctuations in the flue gas flow rate and heat recovery from the WHR, set-point tracking offered by conventional MPC can lead to more energy-intensive operations. \textcolor{black}{Alternatively, economic model predictive control (EMPC), which explicitly incorporates economic operational considerations into a control objective function~\cite{rawlings2012fundamentals, ellis2014tutorial, Wu2022appliedenergy,zhang2019distributed, HAN2024105975, tian2019economic}, holds the promise to ensure safe operations, maintain efficient carbon capture, and minimize energy consumption.}

% \begin{figure*}[t!]
%     \centering
%     \includegraphics[width=1\textwidth]{Fig/structure/framework.eps}
%    % \captionsetup{font={small}}
%     \caption{The proposed hybrid model-based economic model predictive control framework for shipboard post-combustion carbon capture process.}
%     \label{hybrid:fig:entire plant}
% \end{figure*}

In this work, we present a comprehensive shipboard post-combustion carbon capture process design. Considering a practical case scenario when an imperfect first-principles dynamic model is available, we propose a hybrid modeling-based economic model predictive control scheme for efficient process operation. Specifically, we present a complete design for the post-combustion carbon capture process on-board ships which incorporates ship engines, a diesel gas turbine, a waste heat recovery system, and a solvent-based carbon capture plant. Given the dynamic differential algebraic equations representing the first-principles model of this plant may have inaccurately determined parameters, we establish two neural networks; one of them is used as a state dynamics compensator, and the other infers the algebraic states based on the dynamic states and known inputs to the process. The two neural networks are integrated with the imperfect first-principles model to form a high-fidelity hybrid dynamic model for the shipboard post-combustion carbon capture process. An economic model predictive control scheme is developed based on the hybrid model of the shipboard PCC process. The economic cost of the EMPC scheme takes into account fuel expenses and CO$_2$ emission taxes. The effectiveness and performance of the proposed framework are comprehensively evaluated under three ship operational conditions. The proposed approach is also compared with a purely data-driven neural network-based model to illustrate its superiority in terms of data efficiency and model robustness. We presented some preliminary results of this work at the 2024 AIChE Annual Meeting orally, with the abstract available in~\cite{zhang2024learning}.

Contributions of this paper are summarized as follows:

%\vspace{5pt}\parbox[c]{6.3in}{

\begin{enumerate}[label=(\alph*)]
    \vspace{-5pt}\item \textcolor{black}{A ship decarbonization design is formulated to reduce CO$_2$ emissions from ship fuel combustion engines by integrating the post-combustion carbon capture process and seawater cooling with the ship energy generation system.}\vspace{-5pt}

    \item \textcolor{black}{A machine learning-based hybrid modeling method is proposed for the shipboard carbon capture process, with simulation results showcasing good performance in capturing the complex dynamics of the shipboard carbon capture process under varying engine load demands.}\vspace{-5pt}

    \item \textcolor{black}{An economic model predictive control design is proposed based on the developed hybrid model considering the operating cost of the shipboard carbon capture process, with the cross-entropy method~\cite{wen2018constrained,liu2020constrained} being employed to efficiently solve the complex optimization problem involved.}\vspace{-5pt}

    \item \textcolor{black}{The integration of the developed learning-enabled hybrid model and the proposed economic model predictive control streamlines carbon capture process operations, leading to energy-efficient operations with a small impact on carbon capture rate.}
\end{enumerate}
%}

\section{System description and problem formulation}\label{sec:system}

\begin{figure}[t!]
    \centering
    \includegraphics[width=0.485\textwidth]{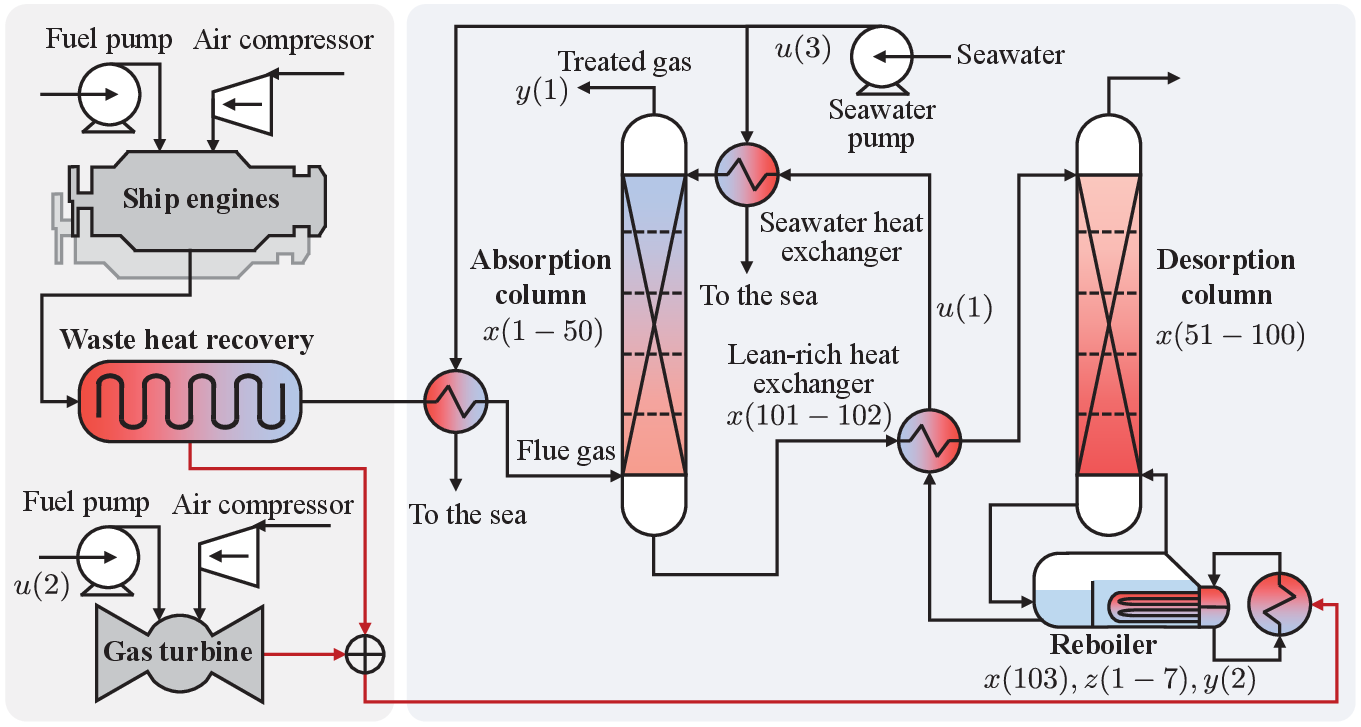}
   % \captionsetup{font={small}}
    \caption{A schematic diagram of the entire plant consisting of the ship engine system and the shipboard post-combustion carbon capture process. The black lines represent mass flow streams, and the red lines represent energy flow streams. $x$, $z$, $u$, $p$, and $y$ denote the differential states, algebraic states, control inputs, known disturbance, and controlled outputs of the entire plant.}\label{hybrid:fig:plant}
\end{figure}

The shipboard carbon capture process comprises two primary systems: the ship engine system and the post-combustion carbon capture plant on-board the ship. A schematic diagram of this integrated process is presented in Figure~\ref{hybrid:fig:plant}. As shown in the gray block of Figure~\ref{hybrid:fig:plant}, the ship engine system encompasses two primary ship engines used for generating propulsion power. The combustion of diesel fuel and air in the two engines generates hot flue gas, which is recovered heat energy in a waste heat recovery (WHR) system. Additional diesel fuel is combusted in a diesel gas turbine to supply extra heat energy. Heat energy from the WHR system and diesel gas turbine is directly supplied to the reboiler. The blue block in Figure~\ref{hybrid:fig:plant} describes the shipboard PCC plant integrated with the ship engine system. In this section, the absorption column is a crucial unit in which the capture of CO$_2$ takes place. In the desorption column, the captured CO$_2$ is released from the solvent by thermal regeneration with heat energy supplied by the reboiler. After regeneration, the hot lean solvent is cooled in a lean-rich solvent and seawater heat exchangers before being fed into the absorption column.

\subsection{Ship engine system}

This work considers a mid-size cargo ship equipped with two 9L46 marine diesel engines~\cite{wartsila46f} for propulsion. The engines are powered by diesel categorized as heavy marine oil. The diesel engines have a fresh air intake and fuel injection systems. Fresh air is pressurized by a turbocharger and is then injected into the engines. The fuel injection system utilizes a fuel pump to control the fuel flow rate into the engines. In the engine, the diesel fuel is mixed with air and then pressurized to a certain pressure, reaching the autoignition temperature. The hot exhaust flue gas from the engines is processed by the WHR system to recover heat energy, and then further cooled and fed into the absorption column for carbon capture. The recovered heat power is utilized to partially energize the reboiler of the shipboard PCC process.

The ship engine load, denoted by $\varphi_E$, can affect the dynamics of the shipboard PCC significantly. During voyages, the operational conditions of ships change, causing fluctuations in engine load. These fluctuations affect the flow rate of flue gas emitted from the engines, which further influences the dynamic operation of the shipboard PCC. Operations such as berthing, cruising, and maneuvering result in varying conditions and capacities for ship engines and PCC plants. These fluctuations introduce uncertainties and make traditional modeling approaches insufficient. In this research, we consider three representative operational conditions: berthing, slow steaming, and maneuvering~\cite{winnes2010emissions}. During berthing, ships remain docked while they load/unload cargo or await their next voyage. Slow steaming, a commonly adopted ship cruising mode, refers to ship operation at sea with a deliberate reduction in the cruising speed. Slow steaming is advantageous for enhancing air pollutant emissions during ship cruising~\cite{tai2022reducing, psaraftis2014ship}. Maneuvering encompasses ship operational activities such as entering/exiting coastal waters, navigating past other vessels, and moving towards or away from a berth or jetty of a port.

\subsubsection{Flue gas flow rate}

The flow rate of CO$_2$ contained in the flue gas, as adapted from~\cite{fan2022review}, is computed as follows:
\begin{equation}\label{eng:co2}
    \tilde{F}_{flue,\text{CO}_2} = \frac{r_{\text{CO}_2}}{3600 r_{\text{C}}} q_{fuel,\text{C}}\varphi_E 2 Q_E W_{SFOC},
\end{equation}
where $\tilde{F}_{flue, \text{CO}_2}$ denotes the mass flow rate of CO$_2$ in the flue gas in kg/s; $r_{\text{C}}$ and $r_{\text{CO}_2}$ are the molar masses of carbon and CO$_2$ in kg/kmol; $q_{fuel, \text{C}}$ denotes the mole fraction of carbon in the fuel; $Q_E$ is the maximum power output from a single engine in kW; $W_{SFOC}$ denotes the specific fuel oil consumption in kg/kWh.

The flue gas flow rate $F_G$ in m$^3$/s can be computed based on the flow rate of CO$_2$, given as follows~\cite{harun2012dynamic}:
\begin{equation}\label{eng:FG}
    F_G = \frac{\tilde{F}_{flue,\text{CO}_2}}{q_{flue,\text{CO}_2}\tilde{\rho}_{flue}},
\end{equation}
where $\tilde{\rho}_{flue}$ is the density of the flue gas in kg/m$^3$; $q_{flue,\text{CO}_2}$ is the mole fraction of CO$_2$ in the flue gas.

\subsubsection{Waste heat recovery of flue gas}

The temperature of the flue gas emitted from the ship engines can reach approximately 360 $^\circ$C. \textcolor{black}{The WHR system cools the flue gas generated by the ship engines to around 150~$^\circ$C to capture the excess heat from the flue gas. The recovered heat energy is then used by the reboiler of the PCC plant.} This way, the energy demand on the diesel gas turbine is reduced. \textcolor{black}{Then, the processed flue gas is further cooled to around 40~$^\circ$C by a heat exchanger and then fed into the absorber for carbon capture.} %The waste heat recovery system captures the excess heat from the flue gas and sends it to the reboiler of the PCC plant.

The amount of the available waste heat energy can be calculated as follows~\cite{jouhara2018waste}:
\begin{equation}\label{heatrec:1}
     Q_{rec} = \tilde{\rho}_{flue}  \tilde{C}_{p,flue} F_{G} (T_{rec,in} - T_{rec,out}),
\end{equation}
where $Q_{rec}$ denotes recovered heat energy in kW; $\tilde{C}_{p,flue}$ is the heat capacity of the flue gas in kJ/(kg$\cdot$K); $T_{rec,in}$ and $T_{rec,out}$ denote the inlet and outlet temperature of the WHR system in K, respectively. The recovered heat energy $Q_{rec}$ will be utilized by the reboiler.

\subsubsection{Diesel gas turbine}

The diesel gas turbine combusts diesel fuel to produce supplementary energy, which is provided to the reboiler of the shipboard PCC plant. The provided heat energy is used to generate hot steam for the reboiler to heat the solvent so that the CO$_2$ can be separated. The heat energy provided by the diesel gas turbine is computed as follows~\cite{danov2004modeling}:
\begin{equation}\label{reb:steam:heat2}
Q_{turbine} = \eta_{fuel} \tilde{F}_{fuel} \frac{\hat{H}_{steam}-\hat{H}_{water}}{\hat{H}_{steam}},
\end{equation}
where $\tilde{F}_{fuel}$ is the mass flow rate of fuel in kg/s; $\eta_{fuel}$ is the calorific value of the diesel fuel in kJ/kg; $Q_{turbine}$ is the heat energy generated by the diesel gas turbine in kW; $\hat{H}_{steam}$ and $\hat{H}_{water}$ are the specific enthalpies of saturated steam and water at a steam pressure of 6 barG in kJ/kg, respectively. The fuel mass flow rate $\tilde{F}_{fuel}$ is a manipulated variable to adjust the reboiler temperature via $Q_{turbine}$.

\textcolor{black}{It is worth noting that the flue gas generated by the diesel gas turbine is not considered and captured by the PCC plant in this work, as its flow rate is relatively small compared to that of the flue gas generated by the ship engines based on simulations.}

\subsection{Shipboard post-combustion carbon capture process}

As shown in Figure~\ref{hybrid:fig:plant}, the shipboard PCC plant comprises five key physical units: the absorption column, the desorption column, the reboiler, the lean-rich solvent heat exchanger, and the seawater heat exchanger. In the absorption column, the lean solvent reacts with the flue gas from the ship engines to selectively absorb CO$_2$ and separate it from other gases in the flue gas stream. The treated gas with a low concentration of CO$_2$ is released into the atmosphere from the top of the absorption column. The rich solvent with absorbed CO$_2$ enters the desorption column for thermal regeneration to release rich CO$_2$ gas at the top. At the bottom of the desorption column, the liquid solvent is transported to the reboiler for heating to approximately 120~$^\circ$C to release and recycle the remaining CO$_2$ back to the desorption column. Meanwhile, the temperature of the hot lean solvent from the reboiler is reduced in the lean-rich heat exchanger and seawater heat exchanger before recycling it back to the top of the absorption column. %\textcolor{black}{The mathematical models of the absorption column, the desorption column, and the lean-rich solvent heat exchanger are adapted from~\cite{decardi2018improving, harun2012dynamic}.} and are omitted in this paper for brevity.

\subsubsection{Absorption column and desorption column}

The absorption column and desorption column have similar model formulations but different reaction directions. Additionally, the desorption column is equipped with a reboiler and a condenser.

%\todo[inline]{----------------delete $\downarrow$----------------}

% \textbf{Modeling assumptions}

% The assumptions made on the absorption and desorption columns include:

% \vspace{5pt}\parbox[c]{6.3in}{

% \begin{enumerate}[(a)]
%     \vspace{-5pt}\item The bulk and liquid phases are well mixed and there are no spacial variations in each stage; \vspace{-20pt}
%     \vspace{-5pt}\item Reactions take place only in the liquid film and the influence of the reaction on mass transfer is characterized using the enhancement factor;
%     \vspace{-5pt}\item Mass and heat transfer are characterized by the two film theory~\cite{whitman1962two};
%     \vspace{-5pt}\item Pressure drop in the absorption and desorption columns is linear;
%     \vspace{-5pt}\item There are no heat losses to the surrounding area.
% \end{enumerate}
% }

% \todo[inline]{----------------delete $\uparrow$----------------}

\textcolor{black}{Based on certain assumptions regarding the operation of the absorption and desorption columns in~\cite{decardi2018improving}, the mass and energy balances in the absorption column and desorption column can be described by the following equations~\cite{decardi2018improving, harun2012dynamic}:}
\begin{align}
    \label{column:balance eq:1} \frac{dC_{L,i}}{dt} &= \frac{4F_L}{\pi D_c^2} \frac{\partial C_{L,i}}{\partial l} + N_i a^I, \\ % \ i = \text{CO}_2, \text{MEA}, \text{H}_2\text{O}, \text{N}_2, \\
    \label{column:balance eq:2} \frac{dC_{G,i}}{dt} &= \frac{4F_G}{\pi D_c^2} \frac{\partial C_{G,i}}{\partial l} - N_i a^I,  \\ % \ i = \text{CO}_2, \text{MEA}, \text{H}_2\text{O}, \text{N}_2, \\
    \label{column:balance eq:3} \frac{dT_L}{dt} &= \frac{4F_L}{\pi D_c^2} \frac{\partial T_L}{\partial l} + \frac{Q_L a^I}{\sum ( C_{L,i} C_{p, i})}, \\ % \ i = \text{CO}_2, \text{MEA}, \text{H}_2\text{O}, \text{N}_2, \\
    \label{column:balance eq:4} \frac{dT_G}{dt} &= \frac{4F_G}{\pi D_c^2} \frac{\partial T_G}{\partial l} + \frac{Q_G a^I}{\sum ( C_{G,i} C_{p, i} )},  % \ i = \text{CO}_2, \text{MEA}, \text{H}_2\text{O}, \text{N}_2,
\end{align}
where $i = \text{CO}_2$, $\text{MEA}$, $\text{H}_2\text{O}$, $\text{N}_2$; $C_{L,i}$ and $C_{G,i}$ denote the liquid and gas phase concentration of component $i$ in kmol/m$^3$; $F_L$ and $F_G$ are the liquid and gas phase volumetric flow rate in m$^3$/s; $D_c$ is the internal diameter of the column in m; $N_i$ is the mass transfer rate in kmol/(m$^2 \cdot$s); $T_L$ and $T_G$ are the liquid and gas temperature in K, respectively; $l$ is the length of the column in m; $C_p$ is the heat capacity in kJ/(kmol$\cdot$K); $Q_L$ and $Q_G$ are the interfacial heat transfer rates for liquid and gas phase in kW/m$^2$, respectively; $a^I$ is the interfacial area in m$^2$/m$^3$. \textcolor{black}{We note that the carbon capture process involves various reactions. In addition to the reaction between the amine and CO$_2$, there are additional amine degradation reactions, which lead to impurities like ammonium salts, carbamate salts, or heat-stable salts~\cite{chao2021post}. %Accurately modeling the dynamic behavior of these reactions using first-principles approaches is typically challenging.
In this study, we primarily focus on the reactions involving MEA for CO$_2$ capture. Moreover, for the purpose of the advanced control approach focused in the present work, the effect of degradation reactions is minor and negligible as a proper solvent management system is assumed.}

\subsubsection{Heat exchanger}
The heat exchangers of the solvent are crucial components of the PCC process. Since there is no transfer of mass between the streams and the assumption of no accumulation, the heat exchanger model does not take mass balances into account. In this shipboard PCC plant, there are two types of heat exchangers: the lean-rich heat exchanger and the seawater heat exchanger. \textcolor{black}{The mathematical model of the lean-rich solvent heat exchanger is adopted from~\cite{decardi2018improving, harun2012dynamic} and is omitted in this paper for brevity. In this subsection, we only present the mathematical model of the seawater heat exchanger.}

The seawater heat exchanger is employed to further decrease the temperature of the hot lean solvent that exits from the lean-rich heat exchanger by utilizing seawater. Seawater serves as an efficient cooling source for reducing the temperature of hot streams in the shipboard PCC process, as it does not involve significant additional expenses. A seawater pump is employed to extract and regulate the seawater flow rate directed to the heat exchanger.

The energy balance of the seawater exchanger, as adapted from~\cite{decardi2018improving, harun2012dynamic}, is presented as follows:
\begin{equation}\label{seahe:T}
    T_{sol, out} = T_{sol, in} + \frac{\hat{\rho}_{sw} F_{sw} \tilde{C}_{p,sw}}{\hat{\rho}_{sol} F_L \tilde{C}_{p,sol}} (T_{sw,in} - T_{sw,out}),
\end{equation}
where $\hat{\rho}$ is the component density in kg/m$^3$; $F_{sw}$ is the seawater volumetric flow rate in m$^3$/s; $\tilde{C}_{p}$ is the heat capacity in kJ/(kg$\cdot$K). Subscripts $sw$ and $sol$ denote seawater and solvent, respectively; $in$ and $out$ denote the inlet and outlet of the seawater heat exchanger, respectively. It is assumed that the density of the seawater and solvent is identical. Seawater volumetric flow rate $F_{sw}$ is one of the manipulated variables to control the solvent temperature.

\subsubsection{Reboiler}

The reboiler is utilized to supply heat to the bottom of the desorption column for the regeneration of the solvent solution. The process of regenerating a CO$_2$ rich amine solution demands substantial energy~\cite{metz2005ipcc}. In this unit, energy is utilized for three primary purposes: (a) breaking chemical bonds between CO$_2$ and MEA; (b) elevating the temperature of the rich solvent solution to its boiling point; (c) vaporizing water to get stripping steam. The solvent containing a low amount of CO$_2$ exits the reboiler as the lean solvent, which is subsequently recycled for further utilization. After passing through the lean-rich heat exchanger and seawater heat exchanger, the lean solvent is reintroduced into the absorption column. To prevent MEA thermal solvent degradation, it is essential to ensure that the reboiler temperature does not surpass 120~$^\circ$C~\cite{alie2006simulation}.

\textbf{Mass balance}

We assume constant liquid levels and pressure in the reboiler, resulting in no vapor or liquid holdup. The mass balance equation of the reboiler is given as follows~\cite{decardi2018improving, harun2012dynamic}:
\begin{equation}\label{reb:mass}\small
    \frac{d M_i}{d t} = \hat{F}_{in} m_{i,in} - \hat{F}_{V} q_{i,out} - \hat{F}_{L} m_{i, out}, \ i = \text{CO}_2, \text{MEA}, \text{H}_2\text{O},
\end{equation}
where $M_i$ denotes the mass holdup of component $i$ in the reboiler in kmol; $\hat{F}_{in}$, $\hat{F}_{V}$, and $\hat{F}_{L}$ denote the inlet stream, vapor, and liquid flow rates in kmol/s, respectively; $m_{i,in}$ and $m_{i,out}$ are the inlet and outlet liquid mole fraction of component $i$ of the reboiler; $q_{i,out}$ is the outlet vapor mole fraction of component $i$ of the reboiler.

\textbf{Energy balance}\label{reb:energy balance}

The dynamic behavior of the reboiler temperature $T_{reb}$ in K can be described based on energy balance with the following form~\cite{decardi2018improving, harun2012dynamic}:
\begin{equation}\label{reb:ene}\small
    \rho_{reb} \hat{C}_{p,reb} V_{reb} \frac{d T_{reb}}{dt} = \hat{F}_{in} H_{L,in} - \hat{F}_{V} H_{V,out} - \hat{F}_{L} H_{L,out} + Q_{reb},
\end{equation}
where $\rho_{reb}$ is the liquid density in the reboiler in kmol/m$^3$; $\hat{C}_{p,reb}$ is the liquid average heat capacity in the reboiler in kJ/(kmol$\cdot$K); $V_{reb}$ is the holdup volume in reboiler in m$^3$; $H_{L,in}$ and $H_{L,out}$ are the liquid enthalpy at the inlet and outlet in kJ/kmol, respectively; $H_{V, out}$ is the vapor enthalpy leaving the reboiler in kJ/kmol; $Q_{reb}$ is the reboiler heat duty in kW.

The reboiler heat duty is derived from the recovered heat energy~(\ref{heatrec:1}) and the diesel gas turbine~(\ref{reb:steam:heat2}), formulated as follows:
\begin{equation}\label{reb:Q}
    Q_{reb} = Q_{rec} + Q_{turbine}.
\end{equation}

\subsection{Compact form of the process model}

The first-principles-based dynamic process model can be written in the form of differential algebraic equations (DAEs) as follows:
\begin{subequations} \label{pcc:model}
 \begin{align}
    \label{pcc:model:1} \dot{x} = f (x, z, u, p), \\
    \label{pcc:model:2} g (x, z, u, p) = 0, \\
    \label{pcc:model:3} y = h (x, p), \quad
\end{align}
\end{subequations}
where $x \in  \mathbb{X} \subset \mathbb{R}^{103}$ denotes the differential states; $z \in \mathbb{Z} \subset \mathbb{R}^7$ denotes the algebraic states; $u \in \mathbb{U} \subset \mathbb{R}^3$ denotes the control inputs; $p \in \mathbb{P} \subset \mathbb{R}^1$ denotes the known disturbance to the system; $y \in \mathbb{Y} \subset \mathbb{R}^2$ denotes the controlled outputs. $\mathbb{X}$, $\mathbb{Z}$, $\mathbb{U}$, $\mathbb{P}$, and $\mathbb{Y}$ are compact sets.

To simulate the process operation, (\ref{pcc:model}) can be discretized using an implicit differential-algebraic solver (IDAS) integrator for DAE systems via CasADi SUNDIALS suite~\cite{andersson2019casadi}. This integrator is a function that takes the state vector at the current sampling time, a set of parameters, and a guess for the algebraic states. It outputs the state vector and algebraic states at the next sampling time. Consequently, a discrete-time expression of~(\ref{pcc:model}) can be obtained as follows:
\begin{subequations}\label{pcc:dis model}
 \begin{align}
    \label{pcc:dis model:1} x_{k+1} = F (&x_k, z_k, u_k, p_k) \\
    \label{pcc:dis model:2} g (x_k, z_k, &u_k, p_k) = 0 \\
    \label{pcc:dis model:3} y_k = h &(x_k, p_k)
\end{align}
\end{subequations}
where $x_k, z_k, u_k$, and $p_k$ denote the differential states, algebraic states, control inputs, and known disturbance at time instant $k$, respectively.

The states of the shipboard PCC process are presented in Figure~\ref{hybrid:fig:plant}. Specifically, the control inputs $u = [F_L, \tilde{F}_{fuel}, F_{sw}]^{\top}$ include the liquid solvent flow rate in m$^3$/s, the flow rate of fuel consumed by the diesel gas turbine in kg/s, and seawater flow rate in m$^3$/s; the known disturbance $p = \varphi_E$ is the ship engine load represented by ratio; the controlled outputs $y = [\tilde{F}_{\text{CO}_2}, T_{reb}]^{\top}$ include the CO$_2$ flow rate in the treated gas in kg/s and reboiler temperature in K. $\tilde{F}_{\text{CO}_2}$ can be computed using the concentration of CO$_2$ in the absorption column and the flue gas flow rate as $\tilde{F}_{\text{CO}_2}= r_{\text{CO}_2} C_{G,\text{CO}_2}^1 F_G$. The carbon capture rate is computed as $\varphi_{\text{CO}_2} = (\tilde{F}_{flue, \text{CO}_2} - \tilde{F}_{\text{CO}_2})/\tilde{F}_{flue, \text{CO}_2}$. The physical meanings of the differential states and algebraic states are listed in Table~\ref{table:diff states} and Table~\ref{table:alg states}, respectively. In Table~\ref{table:diff states}, the $x(i)$ denotes the $i$th variable within the differential states; the superscripts of $C^n$ and $T^n$ denote the concentration and temperature in the $n$th ($n=1,\ldots, 5$) layer of the column. In Table~\ref{table:alg states}, $q_{reb}$ is the vapor fraction of the reboiler; $F_{G,red}$ is the gas phase volumetric flow rate exiting the reboiler in m$^3$/s. It is noted that $z(i)$ denotes the $i$th variables within the algebraic states.

% \begin{figure*}[t!]
%     \centering
%     \includegraphics[width=1\textwidth]{Fig/new/spcc_3.eps}
%    % \captionsetup{font={small}}
%     \caption{An illustration of the state variables of the ship engine system and shipboard PCC plant.}\label{hybrid:fig:plant var}
% \end{figure*}

\newcolumntype{C}[1]{>{\centering\arraybackslash}p{#1}}
\begin{table*}[t!]
    \renewcommand\arraystretch{1.2}
    \caption{Physical meanings of differential states of shipboard PCC process ($n=1,\ldots, 5$).}
    \label{table:diff states}
    \centering
    \begin{tabular}{C{0.1\textwidth}C{0.1\textwidth}|
                    C{0.1\textwidth}C{0.1\textwidth}|
                    C{0.1\textwidth}C{0.1\textwidth}|
                    C{0.1\textwidth}C{0.1\textwidth}}
      \toprule
      \multicolumn{4}{c|}{\textbf{Absorption column}} &
      \multicolumn{4}{c}{\textbf{Desorption column}} \\
      \midrule
      State & {Meaning} &
      State & {Meaning} &
      State & {Meaning} &
      State & {Meaning} \\
      \hline
      $x(1\text{--}5)$ & $C^n_{L,\text{N}_2}$ &
      $x(26\text{--}30)$ & $C^n_{G,\text{N}_2}$ &
      $x(51\text{--}55)$ & $C^n_{L,\text{N}_2}$ &
      $x(76\text{--}80)$ & $C^n_{G,\text{N}_2}$ \\
      $x(6\text{--}10)$ & $C^n_{L,\text{CO}_2}$ &
      $x(31\text{--}35)$ & $C^n_{G,\text{CO}_2}$ &
      $x(56\text{--}60)$ & $C^n_{L,\text{CO}_2}$ &
      $x(81\text{--}85)$ & $C^n_{G,\text{CO}_2}$ \\
      $x(11\text{--}15)$ & $C^n_{L,\text{MEA}}$ &
      $x(36\text{--}40)$ & $C^n_{G,\text{MEA}}$ &
      $x(61\text{--}65)$ & $C^n_{L,\text{MEA}}$ &
      $x(86\text{--}90)$ & $C^n_{G,\text{MEA}}$ \\
      $x(16\text{--}20)$ & $C^n_{L,\text{H}_2\text{O}}$ &
      $x(41\text{--}45)$ & $C^n_{G,\text{H}_2\text{O}}$ &
      $x(66\text{--}70)$ & $C^n_{L,\text{H}_2\text{O}}$ &
      $x(91\text{--}95)$ & $C^n_{G,\text{H}_2\text{O}}$ \\
      $x(21\text{--}25)$ & $T_L^n$ &
      $x(46\text{--}50)$ & $T_G^n$ &
      $x(71\text{--}75)$ & $T_L^n$ &
      $x(96\text{--}100)$ & $T_G^n$ \\
      \midrule
      \multicolumn{4}{c|}{\textbf{Lean-rich solvent heat exchanger}} &
      \multicolumn{4}{c}{\textbf{Reboiler}} \\
      \midrule
      State & {Meaning} &
      State & {Meaning} &
      \multicolumn{2}{c}{State} &
      \multicolumn{2}{c}{Meaning} \\
      \hline
      $x(101)$ & $T_{tube}$ &
      $x(102)$ & $T_{shell}$ &
      \multicolumn{2}{c}{$x(103)$} & \multicolumn{2}{c}{$T_{reb}$} \\
      \bottomrule
    \end{tabular}
  \end{table*}

\begin{table}[t!]
  \renewcommand\arraystretch{1.2}
  \caption{Physical meanings of algebraic states of shipboard PCC process.}\label{table:alg states}
  \centering%\small
    \begin{tabular}{c c | c c}
      \toprule
      % \multicolumn{8}{c}{Algebraic states} \\ \hline
      \multicolumn{4}{c}{Reboiler} \\ \midrule
      State &Meaning &State &Meaning \\ \hline
      $z(1)$ &$C_{L,\text{N}_2}$ &$z(5)$ &$q_{reb}$ \\
      $z(2)$ &$C_{L,\text{CO}_2}$ &$z(6)$ &$m_{\text{CO}_2, out}$ \\
      $z(3)$ &$C_{L,\text{MEA}}$ &$z(7)$ &$F_{G,reb}$ \\
      $z(4)$ &$C_{L,\text{H}_2\text{O}}$ \\ \bottomrule
    \end{tabular}
\end{table}

\subsection{Problem formulation}

From an application point of view, (\ref{pcc:dis model}) is not perfectly known considering the large scale and complex structure of the process. Therefore, we consider a case scenario when only imperfect principles knowledge is available to characterize the dynamics of the process. The direct use of an inaccurate first-principles model can lead to compromised prediction dynamics and process operation.

To compensate for the model mismatch associated with the imperfect first-principles model, machine learning-based hybrid modeling is a promising approach. The objective of this work is twofold: 1) to construct neural networks to infer algebraic state variables and compensate for the model mismatch of the existing imperfect first-principles model, and integrate them seamlessly to form a hybrid model, which can be used to describe the dynamic behavior of the entire shipboard PCC plant; 2) based on the established hybrid model, to develop a computationally efficient economic MPC scheme for efficient and economic operation of the entire shipboard plant in the presence of variations in ship operational conditions.

\section{Hybrid modeling}

In this section, we propose a hybrid modeling approach that incorporates a physics-based component based on the available first-principles knowledge of the entire process and two dense neural networks (DNNs) trained using process data. \textcolor{black}{Hybrid modeling provides the advantages of high prediction accuracy, robust generalization capabilities, and enhanced interpretability~\cite{shah2022deep,bangi2020deep}.}

\begin{figure}[t!]
    \centering
    \includegraphics[width=0.485\textwidth]{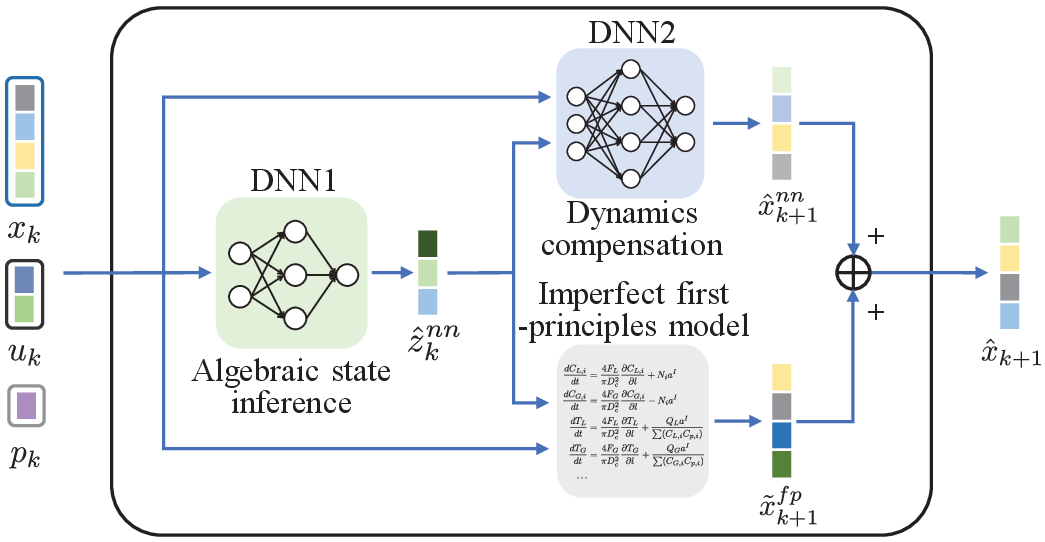}
   % \captionsetup{font={small}}
    \caption{An illustrative diagram of the proposed hybrid model that comprises imperfect first-principles knowledge and dense neural networks. }\label{hybrid:fig:structure}
\end{figure}

The proposed hybrid model is depicted in Figure~\ref{hybrid:fig:structure}. We consider the case where only an imperfect first-principles model is available, which is indicated by the gray block in Figure~\ref{hybrid:fig:structure}. In addition, two dense neural networks are incorporated. The first neural network, which is DNN1 in Figure~\ref{hybrid:fig:structure}, is used to predict the algebraic states based on the differential states, control inputs, and the known system disturbance. The other neural network, which is DNN2 in Figure~\ref{hybrid:fig:structure}, characterizes the discrepancy between the ground-truth of the process and the existing imperfect first-principles dynamic model, by using the inferred algebraic states given by DNN1, the differential states, and control inputs to the process.

\subsection{Imperfect first-principles knowledge}\label{imFP}

The model developed for the shipboard PCC process based on mass and energy balance equations can capture the essential dynamics of the process in general. Due to the complex nature of the process, some mechanisms are not yet understood to represent the process accurately. In such cases, empirical formulations are introduced in the kinetic model. We consider the case where only an imperfect first-principles model is available. The developed shipboard carbon capture model~(\ref{pcc:dis model}) with inaccurate parameters forms the imperfect first-principles model.

Specifically, the mass transfer coefficients of gas and liquid phases of different components, the interfacial heat transfer coefficient, and the enhancement factor for the desorption column are inaccurate.  The mass transfer coefficients in gas and liquid phases are used to compute the mass transfer rate. The mass transfer coefficients are determined from generalized correlations proposed in~\cite{onda1968mass}. The interfacial heat transfer coefficient is computed using the Chilton-Colburn analogy given in~\cite{geankoplis2003transport}. The enhancement factor for the desorption column is computed based on~\cite{tobiesen2008experimental}. The constant coefficients of the empirical formulations for the aforementioned parameters are slightly changed to represent imperfect first-principles knowledge.

Based on the developed shipboard PCC model~(\ref{pcc:dis model}), the imperfect first-principles model of the shipboard PCC process is denoted as follows:
\begin{equation}\label{pcc:fp}
     \tilde{x}^{fp}_{k+1} = F_{fp}(x_k, z_k, u_k, p_k),
\end{equation}
where $\tilde{x}^{fp}_{k+1}$ denotes the differential states computed by the imperfect first-principles model at time instant~$k+1$.

\subsection{Algebraic state inference}

The underlying first-principles model for the entire process is described by a set of DAEs. Due to the existence of algebraic equations, solving DAE problems can be much more computationally complex than solving ordinary differential equations (ODEs). \textcolor{black}{Based on these insights, we construct a neural network to infer the algebraic states at each time instant. According to~(\ref{pcc:dis model:2}), the algebraic states~$z$ can be represented as a function of the differential states~$x$, control inputs~$u$, and known disturbance~$p$. This leads to the formulation of the algebraic state inference neural network in the following form:}
\begin{equation}\label{nn:z:model}
    \hat{z}^{nn}_k = G_{nn} \big( x_k, u_k, p_k | \theta_{G_{nn}} \big),
\end{equation}
where $\hat{z}^{nn}_k$ denotes the inferred algebraic states at time instant $k$; $\theta_{G_{nn}}$ denotes the parameters of this neural network. \textcolor{black}{The neural network $G_{nn}$ in~(\ref{nn:z:model}) is constructed as a fully-connected feedforward neural network~\cite{bebis1994feed}. Other neural network structures such as long-short-term memory (LSTM)~\cite{yu2019review} may also be adopted.}

Given a data set~$\mathcal{D}_z$, the neural network in~(\ref{nn:z:model}) is trained to minimize the discrepancy between the ground-truth of the algebraic states and the inferred values. Therefore, the optimization problem can be formulated as follows:
\begin{subequations}\label{nn:z:opt}
    \begin{align}
        \label{nn:z:opt:1} \min_{\theta_{G_{nn}}} \ &\mathcal{L}_z (\theta_{G_{nn}}) = \mathbb{E}_{\mathcal{D}_z} \left \Vert z_k - \hat{z}^{nn}_k \right \Vert^2_2 \\
        \label{nn:z:opt:3}  \text{s.t.} \quad &\hat{z}^{nn}_k = G_{nn} \big( x_k, u_k, p_k | \theta_{G_{nn}} \big)
    \end{align}
\end{subequations}

\subsection{State dynamics compensation}

\textcolor{black}{To account for the mismatch between the real plant in~(\ref{pcc:dis model:1}) and the existing imperfect first-principles model in~(\ref{pcc:fp}), a state dynamics compensation neural network is trained and incorporated into the proposed hybrid model. According to the differential equations described in~(\ref{pcc:dis model:1}) and~(\ref{pcc:fp}), the differential states~$x_{k+1}$ for the next time instant~$k+1$ is only affected by the current states~$x_k$, $z_k$, $u_k$, and~$p_k$ at time instant~$k$. Based on this observation, the state dynamics compensation neural network used for predicting the differential state error is formulated as follows:}
\begin{equation}\label{nn:x:model}
    \hat{x}^{nn}_{k+1} = F_{nn} \big(x_k, z_k, u_k, p_k | \theta_{F_{nn}}\big),
\end{equation}
where $\hat{x}^{nn}_{k+1}$ denotes the predicted differential state error at time instant~$k+1$; $\theta_{F_{nn}}$ denotes the parameters of the neural network to be trained using process data. \textcolor{black}{Similar to the algebraic state inference neural network, $F_{nn}$ in~(\ref{nn:x:model}) is constructed as a fully-connected feedforward neural network~\cite{bebis1994feed}. It may also be constructed based on other neural network structures, such as LSTM~\cite{yu2019review}.} Based on the existing imperfect first-principles model in~(\ref{pcc:fp}), a prediction of the differential states, denoted by~$\hat{x}_{k+1}$, is obtained as follows:
\begin{equation}\label{x:fp+nn}
    \hat{x}_{k+1} = \tilde{x}^{fp}_{k+1} + \hat{x}^{nn}_{k+1}.
\end{equation}

The objective of introducing this dynamics compensation neural network is to minimize the mismatch between the prediction of the future differential states and the ground-truth data. Therefore, given a data set~$\mathcal{D}_x$, the objective function used to train the parameters~$\theta_{F_{nn}}$ is:
\begin{subequations}\label{nn:x:loss}
    \begin{align}
        \label{eq:sdc:obj:1} \mathcal{L}_x (\theta_{F_{nn}}) &= \mathbb{E}_{\mathcal{D}_x} \left \Vert x_{k+1} - \hat{x}_{k+1} \right \Vert_2^2, \\
        \label{eq:sdc:obj:2} &= \mathbb{E}_{\mathcal{D}_x} \left \Vert x_{k+1} - \big(\tilde{x}^{fp}_{k+1} + \hat{x}^{nn}_{k+1}\big) \right \Vert_2^2, \\
         \label{eq:sdc:obj:3} &= \mathbb{E}_{\mathcal{D}_x} \left  \Vert x^e_{k+1} - \hat{x}^{nn}_{k+1} \right \Vert_2^2 ,
    \end{align}
\end{subequations}
\textcolor{black}{where $x_{k+1}$ denotes the real differential states; $x^e_{k+1} = x_{k+1} - \tilde{x}^{fp}_{k+1}$ denotes the mismatch between the prediction of the imperfect first-principles model and the ground-truth data. It is worth noting that the ground-truth data is generated using the comprehensive first-principles model in~(\ref{pcc:dis model}), and is not real experimental data from a shipboard PCC plant.}

The optimization problem for the neural network training can be formulated as follows:
\begin{subequations}\normalsize
    \begin{align}
        &\min_{\theta_{F_{nn}}} \ \mathcal{L}_x  (\theta_{F_{nn}}) \\
        % \text{s.t.} \ &\mathcal{L}_x (\theta_{F_{nn}}) = \mathbb{E}_{\mathcal{D}_x} \Big|\Big| \big(x_k - x^{fp}_k\big) - x^{nn}_k\Big|\Big|_2^2 \\
        \text{s.t.} \quad &\hat{x}_{k+1} = \tilde{x}^{fp}_{k+1} + \hat{x}^{nn}_{k+1} \\
        &\hat{x}^{nn}_{k+1} = F_{nn} \big(x_k, z_k, u_k, p_k | \theta_{F_{nn}}\big) \\
        & \tilde{x}^{fp}_{k+1} = F_{fp}(x_k, z_k, u_k, p_k)
    \end{align}
\end{subequations}

\begin{rmk}
    \textcolor{black}{Hybrid modeling that leverages neural networks to compensate for the mismatch between the first-principles knowledge and the actual process dynamics has been explored, see, e.g., in~\cite{oliveira2004combining, su1993integrating}, with applications to a bioreactor system~\cite{oliveira2004combining} and polymer reaction~\cite{su1993integrating}. In the current work, we also leverage machine learning to capture and compensate for the discrepancies between the first-principles model and the plant. In particular, to address the challenge of handling the first-principles model in the form of DAE, the algebraic state inference network in~(\ref{nn:z:model}), which serves as a non-parametric static equation, is introduced to infer the algebraic states, which facilitates the establishment of the machine learning-based hybrid dynamic model for shipboard PCC.}
\end{rmk}

\begin{rmk}
    \textcolor{black}{Although the proposed hybrid model includes a state dynamics compensation neural network to reduce the model mismatch, it cannot completely eliminate the modeling errors, and does not fully account for the unmeasured disturbances in real applications. Online model updates through feedback correction using real-time information can be helpful for further reducing model error. For example, Levenberg-Marquardt (LM) algorithm may be utilized to update the neural network parameters~\cite{vatankhah2019nonlinear}. During online implementation, the parameters of the neural networks can be trained using new data by minimizing an objective function, with the LM algorithm. With an even more accurate model, control performance may be further improved. However, continuously updating the neural network during online implementation may result in overfitting and inefficient computation~\cite{hong2019fault}. How to develop a comprehensive online correction strategy for the proposed hybrid model will be considered in future research.}
\end{rmk}

\begin{figure}[t!]
    \centering
    \includegraphics[width=0.485\textwidth]{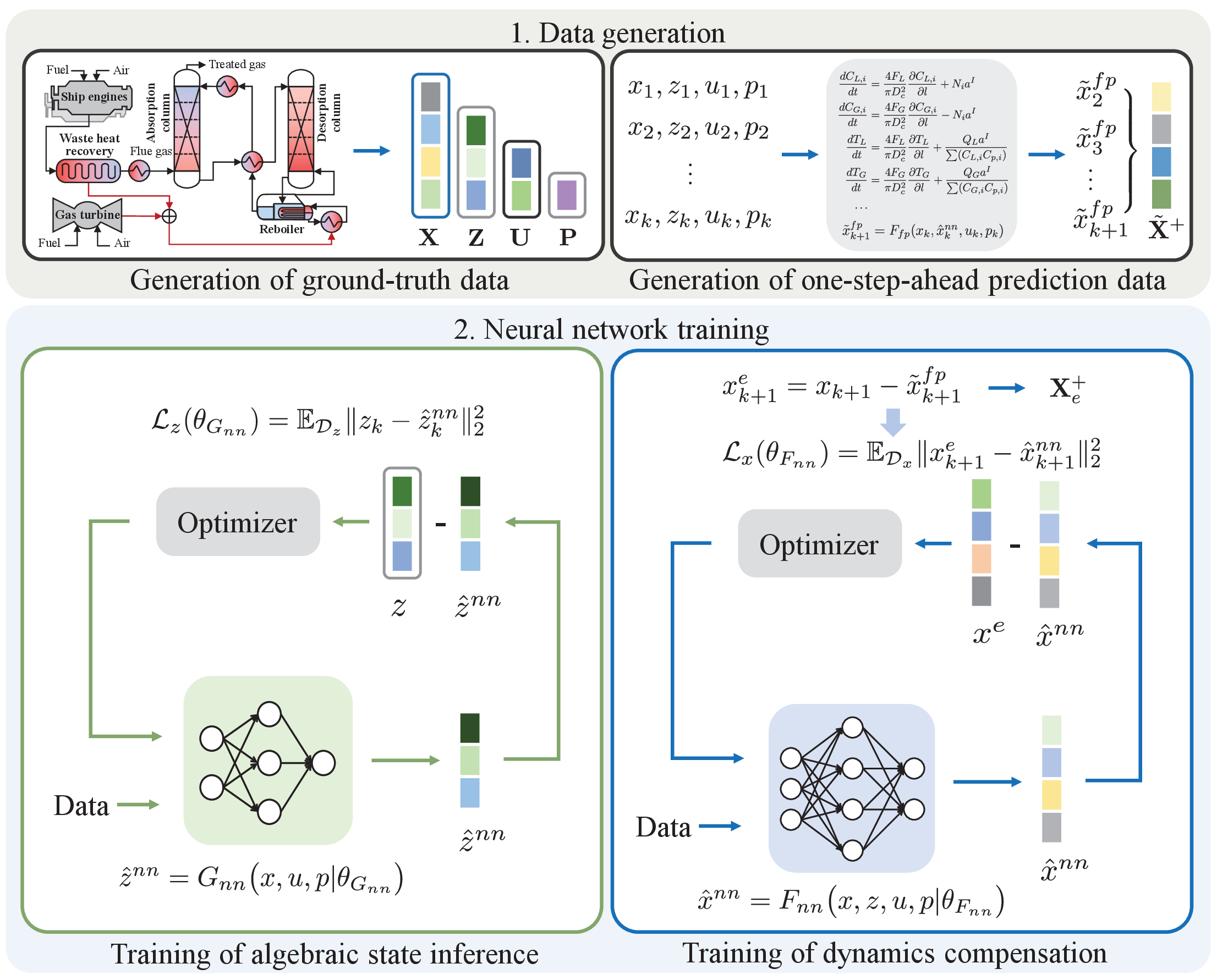}
   % \captionsetup{font={small}}
    \caption{A block diagram illustrating the training process of the hybrid model.}\label{hybrid:fig:train pred}
\end{figure}

\subsection{The proposed hybrid model and training process}

The full expression of the hybrid dynamic model incorporating the imperfect first-principles model and learned parameters~$\theta_{G_{nn}}^*$ and~$\theta_{F_{nn}}^*$ is given as follows:
\begin{align}\label{hybrid model: x}
    \hat{x}_{k+1} &= F_{fp} \Big(x_k, G_{nn}\big(x_k, u_k, p_k|\theta_{G_{nn}}^*\big) , u_k, p_k \Big) \nonumber \\  
    &\ +  F_{nn} \Big(x_k, G_{nn}\big(x_k, u_k, p_k|\theta_{G_{nn}}^*\big) , u_k, p_k|\theta_{F_{nn}}^* \Big) .
\end{align}

The hybrid model training and the implementation of the hybrid model for state prediction are illustrated in Figure~\ref{hybrid:fig:train pred}. \textcolor{black}{The first block in Figure~\ref{hybrid:fig:train pred} indicates that the training of the algebraic state inference network and state dynamics compensation network requires simulation data obtained from the comprehensive first-principles model and the imperfect first-principles model. The comprehensive first-principles model of the shipboard PCC plant in~(\ref{pcc:dis model}) is used as a high-fidelity simulator to generate ground-truth data. As illustrated in the second block in Figure~\ref{hybrid:fig:train pred}, the algebraic state inference and state dynamics compensation networks are trained based on their respective loss functions. After training is completed, the established hybrid model can be used to predict future process states.}% based on real-time information from the ship engine system and the associated PCC plant, and to develop advanced control schemes, as shown in the third block of Figure~\ref{hybrid:fig:train pred}.}

\begin{algorithm}[t!]
\vspace{1.5mm}\small
\caption{\small Training the algebraic state inference}\label{alg:train}
\KwIn{The number of neural network layers, the number of neurons; the number of epoch $N_{epochs,z}$; learning rate; open-loop training data $\mathbf{X}, \mathbf{U}, \mathbf{P}$, and label data $\mathbf{Z}$ using shipboard carbon capture process simulator in~(\ref{pcc:dis model}).}  \vspace{2.5mm}

\KwOut{Algebraic state inference neural network with optimized parameter $\theta_{G_{nn}}^*$.} \vspace{2.5mm}

\SetAlgoLined

\parbox[t]{0.9\linewidth}{
    \For{$n = 1, \ldots, N_{epochs, z}$}{\vspace{2.5mm}
        \begin{enumerate}[label=\textbf{\scriptsize 1.\arabic*}]
            \item Compute inferred algebraic states: \vspace{1mm} \\ 
            $ \hat{z}^{nn}_k = G_{nn} \big( x_k, u_k, p_k | \theta_{G_{nn}} \big)$.

            \item Compute the algebraic loss: \vspace{1mm} \\ 
            $\mathcal{L}_z (\theta_{G_{nn}}) = \mathbb{E}_{\mathcal{D}_z} \left \Vert z_k - \hat{z}^{nn}_k \right \Vert^2_2$.

            \item Update parameter $\theta_{G_{nn}}$ by gradient descent.
        \end{enumerate}\vspace{-3mm}
    }
}
\end{algorithm}

% \begin{algorithm*}[t!]
% \vspace{1.5mm}
% \caption{Training the algebraic state inference}\label{alg:train}
% \KwIn{The number of neural network layers, the number of neurons; the number of epoch $N_{epochs,z}$; learning rate; open-loop training data $\mathbf{X}, \mathbf{U}, \mathbf{P}$, and label data $\mathbf{Z}$ using shipboard carbon capture process simulator in~(\ref{pcc:dis model}).}  \vspace{2.5mm}

% \KwOut{Algebraic state inference neural network with optimized parameter $\theta_{G_{nn}}^*$.} \vspace{2.5mm}

% \SetAlgoLined

% \parbox[t]{0.9\linewidth}{
%     \For{$n = 1, \ldots, N_{epochs, z}$}{\vspace{2.5mm}
%         \begin{enumerate}[label=\textbf{\scriptsize 1.\arabic*}]
%             \item Compute inferred algebraic states: $ \hat{z}^{nn}_k = G_{nn} ( x_k, u_k, p_k | \theta_{G_{nn}} )$.

%             \item Compute the algebraic loss: $\mathcal{L}_z (\theta_{G_{nn}}) = \mathbb{E}_{\mathcal{D}_z} || z_k - \hat{z}^{nn}_k||^2_2$.

%             \item Update the parameter $\theta_{G_{nn}}$ by gradient descent.
%         \end{enumerate}\vspace{-3mm}
%     }
% }
% \end{algorithm*}

\begin{algorithm}[t!]
    \vspace{1.5mm}\small
    \caption{\small Training the state dynamics compensation}\label{state:train}
    \KwIn{The number of neural network layers, the number of neurons; the number of epoch $N_{epochs,x}$; learning rate; open-loop training data $\mathbf{X}, \mathbf{Z}, \mathbf{U}, \mathbf{P}$, and label data $\mathbf{X}^+$ using shipboard carbon capture process simulator in~(\ref{pcc:dis model}); one-step-ahead prediction data $\tilde{\mathbf{X}}^+$ based on imperfect first-principles model in~(\ref{pcc:fp}).}  \vspace{2.5mm}
    
    \KwOut{State dynamics compensation neural network with optimized parameter $\theta^*_{F_{nn}}$.} \vspace{2.5mm}
    
    \SetAlgoLined
    
    \textcolor{black}{Compute the mismatch label data $\mathbf{X}^+_e = \mathbf{X}^+ - \tilde{\mathbf{X}}^+$.}\vspace{2.5mm}
    
    \parbox[t]{0.9\linewidth}{
        \For{$n = 1, \ldots, N_{epochs, x}$}{\vspace{2.5mm}
            \begin{enumerate}[label=\textbf{\scriptsize 1.\arabic*}]
                \item Compute one-step-ahead state compensation: \vspace{1mm} \\
                 $\hat{x}^{nn}_{k+1} = F_{nn} \big(x_k, z_k, u_k, p_k | \theta_{F_{nn}} \big)$.
    
                \item \textcolor{black}{Compute the differential state loss: \vspace{1mm} \\
                 $\mathcal{L}_x (\theta_{F_{nn}}) = \mathbb{E}_{\mathcal{D}_x} \left \Vert x^e_{k+1} - \hat{x}^{nn}_{k+1} \right \Vert_2^2$.}
    
                \item Update parameter $\theta_{F_{nn}}$ by gradient descent.
            \end{enumerate}\vspace{-3mm}
        }
    }
    \end{algorithm}

% \begin{algorithm*}[t!]
% \vspace{1.5mm}
% \caption{Training the state dynamics compensation}\label{state:train}
% \KwIn{The number of neural network layers, the number of neurons; the number of epoch $N_{epochs,x}$; learning rate; open-loop training data $\mathbf{X}, \mathbf{Z}, \mathbf{U}, \mathbf{P}$, and label data $\mathbf{X}^+$ using shipboard carbon capture process simulator in~(\ref{pcc:dis model}); one-step-ahead prediction data $\tilde{\mathbf{X}}^+$ based on imperfect first-principles model in~(\ref{pcc:fp}).}  \vspace{2.5mm}

% \KwOut{State dynamics compensation neural network with optimized parameter $\theta^*_{F_{nn}}$.} \vspace{2.5mm}

% \SetAlgoLined

% \textcolor{black}{Compute the mismatch label data $\mathbf{X}^+_e = \mathbf{X}^+ - \tilde{\mathbf{X}}^+$.}\vspace{2.5mm}

% \parbox[t]{0.9\linewidth}{
%     \For{$n = 1, \ldots, N_{epochs, x}$}{\vspace{2.5mm}
%         \begin{enumerate}[label=\textbf{\scriptsize 1.\arabic*}]
%             \item Compute one-step-ahead state compensation: $\hat{x}^{nn}_{k+1} = F_{nn} (x_k, z_k, u_k, p_k | \theta_{F_{nn}})$.

%             \item \textcolor{black}{Compute the differential state loss: $\mathcal{L}_x (\theta_{F_{nn}}) = \mathbb{E}_{\mathcal{D}_x} || x^e_{k+1} - \hat{x}^{nn}_{k+1}||_2^2$.}

%             \item Update the parameter $\theta_{F_{nn}}$ by gradient descent.
%         \end{enumerate}\vspace{-3mm}
%     }
% }
% \end{algorithm*}

%The training process of the hybrid model is summarized as Algorithm~\ref{alg:train}. The training of the algebraic state inference and state dynamics compensation are separated.
The training processes of the algebraic state inference and state dynamics compensation are summarized as Algorithm~\ref{alg:train} and Algorithm~\ref{state:train}, respectively. \textcolor{black}{It is worth mentioning that, to compute the objective function~(\ref{eq:sdc:obj:3}) effectively during the training process of the neural network, the mismatch for the differential states between the prediction of the imperfect first-principles model and the ground-truth data, $x^e_{k+1}$, shall be constructed first. Specifically, based on the given open-loop data set~$\mathcal{D}_x$ that consists of ground-truth data~$\mathbf{X}$, $\mathbf{Z}$, $\mathbf{U}$, and $\mathbf{P}$ generated by simulating the comprehensive first-principles model~(\ref{pcc:dis model}), the data of one-step-ahead prediction for the differential states, denoted by~$\tilde{\mathbf{X}}^+$, can be obtained by simulating the imperfect first-principles model in~(\ref{pcc:fp}) for one step. The data of the mismatch for the differential states, denoted by~$\mathbf{X}^+_e$, can then be obtained by subtracting the one-step-ahead prediction from the actual differential states at the same time instant.} To predict the future dynamics of the shipboard carbon capture process, the hybrid model only requires the current differential states, control inputs, and the known disturbance, while the algebraic state information is not needed.

\begin{rmk}
    \textcolor{black}{In real applications, measuring certain state variables of the shipboard PCC plant in real-time using hardware sensors can be challenging or expensive. The current study primarily focuses on the hybrid modeling of the shipboard PCC plant and the application of economic model predictive control based on the proposed hybrid model. To enhance practical applications, state estimation can be incorporated to reconstruct the information of all the system states without deploying additional hardware sensors. Specifically, state estimation leverages a dynamic model to develop a state estimator, which can be used to provide real-time estimates of all the system states, including the unmeasured states, based on the available sensor measurement information~\cite{muske1995nonlinear, alsac1998generalized, yin2017distributed}. Integrating state estimation methods with the hybrid model-based EMPC approach can enhance the practicality of advanced control for shipboard PCC plants. This direction may be explored in future research.}
\end{rmk}

\section{Economic MPC based on the hybrid model}

In this section, we develop an economic model predictive control (EMPC) scheme based on the developed hybrid model. Instead of attempting to force the process operation towards an optimal steady-state point, EMPC incorporates a general economic cost function into dynamic optimization, and is more promising for the energy-efficient operation of the shipboard carbon capture plant.

\subsection{Economic cost function}

The economic operation is designed to achieve two primary objectives: 1) to mitigate the release of CO$_2$ into the atmosphere, and 2) to minimize the energy consumption needed for process operation. In the proposed EMPC design, we use an objective function that explicitly incorporates both goals mentioned above.

The economic cost function comprises two components: the carbon tax applied to the released CO$_2$ and the cost associated with the energy consumption, and is expressed as follows:
\begin{equation}\label{empc:cost}
     \ell (y_k, u_k) = \alpha \max\big (y_{k}(1) - y_{limit},0 \big) + \beta u_{k}(2),
\end{equation}
where $\alpha$ indicates the carbon tax in $\$$/kg; $\beta$ indicates the price for the fuel in $\$$/kg; $\max \big (y_{k}(1) - y_{limit},0 \big )$ represents the quantity of CO$_2$ emitted beyond the CO$_2$ release threshold at time instant $k$ in kg/s; $y_{limit}$ is the CO$_2$ release threshold in kg/s, beyond which costs will be accrued; $u_k(2)$ and $y_k(1)$ indicate the second variable of the control inputs and first variable of the controlled outputs, respectively. \textcolor{black}{In~(\ref{empc:cost}), the `$\max$' function compares $y_k(1) - y_{limit}$ and $0$, and returns the larger one. The purpose of introducing the `$\max$' function is to ensure that the accrued cost for released CO$_2$ is non-negative.}

\subsection{Optimization problem}

Based on the hybrid model in~(\ref{hybrid model: x}) and the economic cost function in~(\ref{empc:cost}), the optimization problem for the proposed EMPC scheme at time instant~$k$ can be formulated as follows:
\begin{subequations}\label{empc:opt}\small
    \begin{align}
        % \label{empc:opt:min} \min_{u_{k|k}, \ldots, u_{k+N_c-1|k}} \quad & \mathcal{J}_k = \sum_{j=k}^{k+N_c-1} \big( \alpha u_{j|k}(2) + \beta \max(\hat{y}_{j|k}(1) - y_{limit},0) \big)  \\
        %\label{empc:opt:min} \min_{u_{k|k}, \ldots, u_{k+N_c-1|k}} \quad & \mathcal{J}_k = \sum_{j=k}^{k+N_c-1} \big( \alpha u_{j|k}(2) + \beta \epsilon_{j|k} \big)  \\
        \label{empc:opt:min} &\min_{u_{k|k}, \ldots, u_{k+N_c-1|k}} \  \mathcal{J}_k = \sum_{j=k}^{k+N_c-1} \ell (\hat{y}_{j|k}, u_{j|k})  \\
        \label{empc:opt:1} \text{s.t.} \ &\hat{x}_{j+1|k} = F_{fp} \Big (\hat{x}_{j|k}, G_{nn}\big(\hat{x}_{j|k}, u_{j|k}, p_{j|k}|\theta_{G_{nn}}^*\big), u_{j|k}, p_{j|k} \Big ) \nonumber \\
        &\quad + F_{nn} \Big(\hat{x}_{j|k}, G_{nn}\big(\hat{x}_{j|k},u_{j|k}, p_{j|k}|\theta_{G_{nn}}^*\big), u_{j|k}, p_{j|k}|\theta_{F_{nn}}^* \Big),  \\
        \label{empc:opt:2} &\hat{x}_{k|k}=x_k, \\
        \label{empc:opt:3} &\hat{y}_{j|k}=h\big(\hat{x}_{j|k}, p_{j|k}\big) \in \mathbb{Y}, \\
        \label{empc:opt:4} &u_{j|k} \in \mathbb{U}, \quad j = k, \ldots, k+N_c-1,
    \end{align}
\end{subequations}
where $u_{k|k}, \ldots, u_{k+N_c-1|k}$ is a sequence of control inputs at given time instant~$k$; $N_c$ is the control horizon; $\mathcal{J}_k$ is the accumulated cost across~$N_c$ steps at time instant~$k$. In the optimization problem, (\ref{empc:opt:1}) represents the hybrid model of the shipboard carbon capture process; (\ref{empc:opt:2}) is the initial condition of the system at time instant~$k$; and (\ref{empc:opt:3})-(\ref{empc:opt:4}) are the constraints of the controlled outputs and control inputs, respectively.

The optimal solution to the optimization problem~(\ref{empc:opt}) at time instant~$k$ is a sequence of control inputs. The first control action in the optimal control sequence, denoted by~$u^*_{k|k}$, is applied to the shipboard PCC process to achieve the desired control objectives.

\subsection{Cross-entropy method for optimization}

Because of the complex structure and nonlinearity of the hybrid model, applying a gradient-based optimizer to the optimization problem~(\ref{empc:opt}) can be challenging and extremely time-consuming. An alternative method to solve the aforementioned optimization problem is the cross-entropy (CE) method~\cite{de2005tutorial}. The CE method is a state-of-the-art approach for addressing complex optimization problems and has been used in many works~\cite{liu2020constrained, bharadhwaj2020model, wen2018constrained}. Random trajectories are sampled from the distribution repeatedly, and the parameters of the sampling distribution are updated recursively. Through this iterative process, the random trajectories, representing potential solutions to the optimization problem, recursively converge toward the optimal or near-optimal solution. In this study, we employ the constrained CE method~\cite{wen2018constrained,liu2020constrained}, which can tackle the optimization problem with constraints.

% \begin{figure*}[!ht]
%     \centering
%     \captionsetup[subfigure]{font=small}

%     \subfigure[Start with initial distribution.]{
%     \label{fig:cem:1}
%     \includegraphics[width=0.3\textwidth]{Fig/cem/cem_fig1.eps}
%     %\captionsetup{font={small}}
%     }
%     \hfill
%     \subfigure[Sample $N_{sample}$ control sequences.]{
%     \label{fig:cem:2}
%     \includegraphics[width=0.3\textwidth]{Fig/cem/cem_fig2.eps}
%     }
%     \hfill
%     \subfigure[Select the feasible sample set $\Omega$.]{
%     \label{fig:cem:3}
%     \includegraphics[width=0.31\textwidth]{Fig/cem/cem_fig3.eps}
%     }

%     \subfigure[Select the elite sample set $\Lambda$.]{
%     \label{fig:cem:4}
%     \includegraphics[width=0.3\textwidth]{Fig/cem/cem_fig4.eps}
%     %\captionsetup{font={small}}
%     }
%     \hfill
%     \subfigure[Update sampling distribution.]{
%     \label{fig:cem:5}
%     \includegraphics[width=0.3\textwidth]{Fig/cem/cem_fig5.eps}
%     }
%     \hfill
%     \subfigure[Repeat with updated distribution.]{
%     \label{fig:cem:6}
%     \includegraphics[width=0.3\textwidth]{Fig/cem/cem_fig6.eps}
%     }
%     \caption{A graphical illustration of the recursively updating process of the cross-entropy method.}
%     \label{fig:cem}
% \end{figure*}

\begin{figure}[!t]
    \centering
    \captionsetup[subfigure]{font=small}

    \subfigure[Start with initial distribution.]{
    \label{fig:cem:1}
    \includegraphics[width=0.22\textwidth]{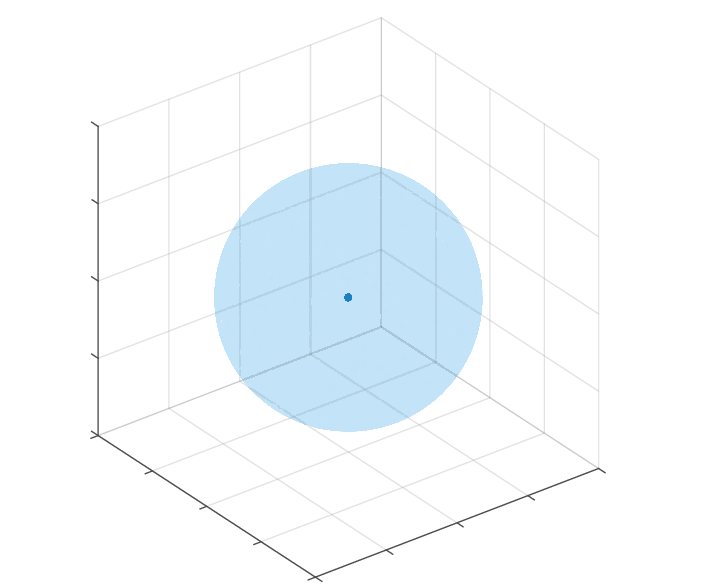}
    %\captionsetup{font={small}}
    }
    \hfill
    \subfigure[Sample $N_{sample}$ control sequences.]{
    \label{fig:cem:2}
    \includegraphics[width=0.22\textwidth]{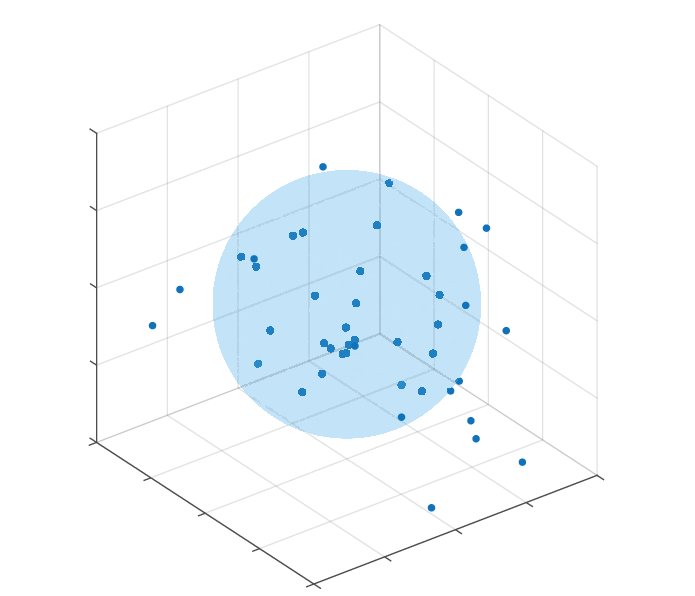}
    }

    \subfigure[Select the feasible sample set $\Omega$.]{
    \label{fig:cem:3}
    \includegraphics[width=0.22\textwidth]{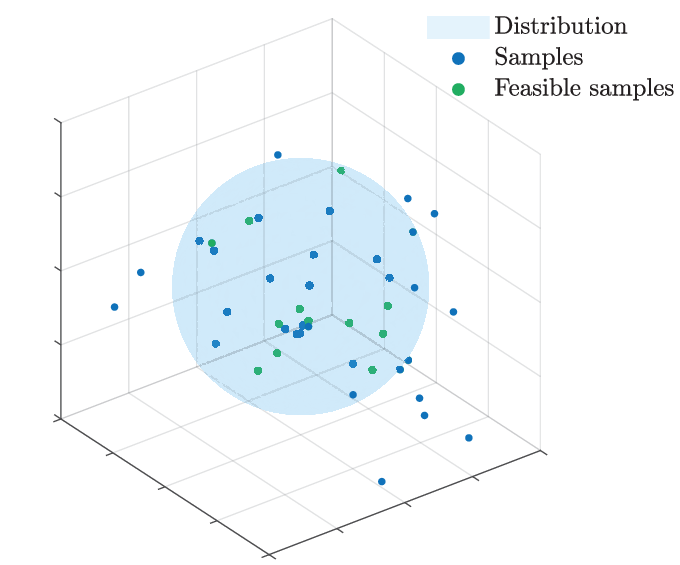}
    }
    \hfill
    \subfigure[Select the elite sample set $\Lambda$.]{
    \label{fig:cem:4}
    \includegraphics[width=0.22\textwidth]{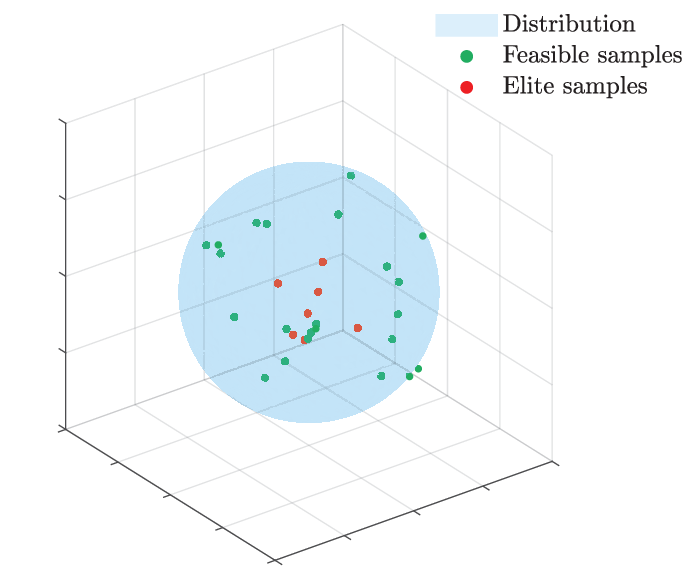}
    %\captionsetup{font={small}}
    }

    \subfigure[Update sampling distribution.]{
    \label{fig:cem:5}
    \includegraphics[width=0.22\textwidth]{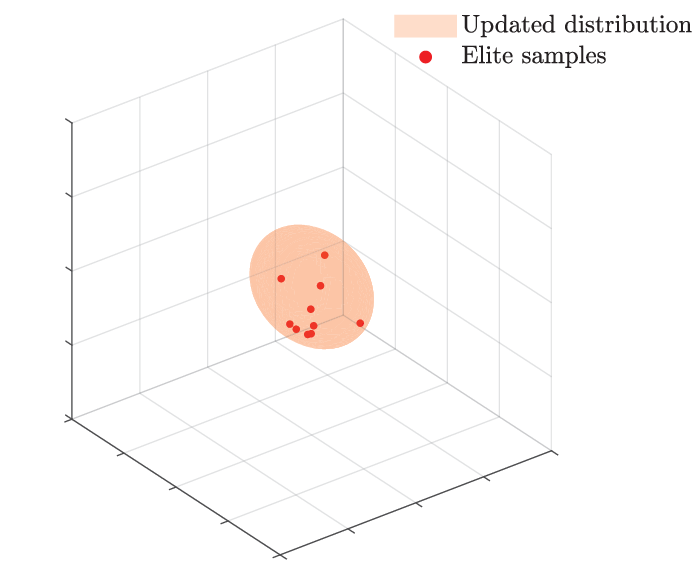}
    }
    \hfill
    \subfigure[Repeat with updated distribution.]{
    \label{fig:cem:6}
    \includegraphics[width=0.22\textwidth]{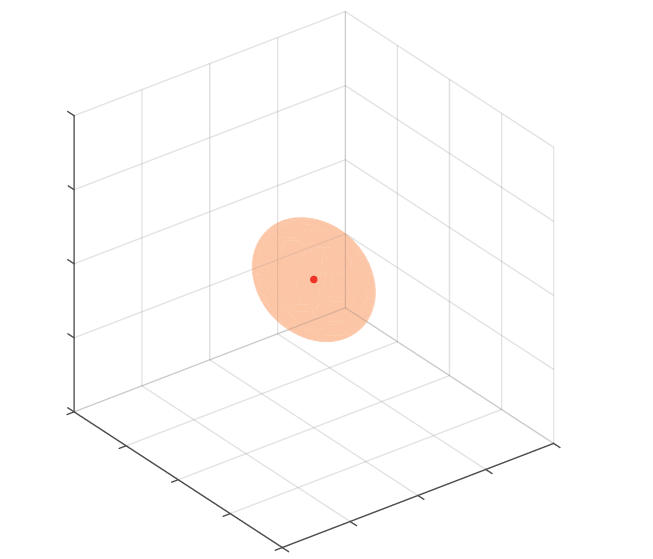}
    }
    \caption{A graphical illustration of the recursively updating process of the cross-entropy method.}
    \label{fig:cem}
\end{figure}

\begin{algorithm*}[t!]
\vspace{1.5mm}\small
\caption{\small One-time execution of EMPC using the CE method.}\label{CEM:MPC}
\KwIn{Iteration number~$N_{CE}$; sample number per iteration~$N_{sample}$; control horizon~$N_c$; size of the elite samples~$N_K$; a minimum variance bound~$\nu_{min}$, update rate~$\lambda$; initial sampling distribution~$\mathcal{N}\big(\mu^0_{k:k+N_c-1}, \nu^0_{k:k+N_c-1}\big)$, initial state~$\hat{x}_{k|k} = x_k$.}  \vspace{1.5mm}

\KwOut{Optimal control action $u^*_k$ at time instant $k$.} \vspace{1.5mm}

\SetAlgoLined

Set $i=0$. \vspace{1.5mm}

\parbox[t]{0.92\linewidth}{\While{$i < N_{CE}$ and $\max\big (\nu^i_{k:k+N_c-1} \big ) > \nu_{min}$}{
    \vspace{1.5mm}
    \begin{enumerate}[label=\textbf{\scriptsize 2.\arabic*}]
        \item Sample $\Big\{ \big(u_{k|k}^i, \ldots, u_{k+N_c-1|k}^i \big)_m \Big\}_{m=1}^{N_{sample}}  \sim \mathcal{N}\big(\mu^i_{k:k+N_c-1}, \nu^i_{k:k+N_c-1}\big)$ and clip the control inputs according to~(\ref{empc:opt:4}).

        \item Predict the future states and controlled outputs according to~(\ref{empc:opt:1}) and~(\ref{empc:opt:3}), and the objective function values.

        \item  Select the feasible sample set $\Omega$ of which controlled outputs satisfy the output constraints according to~(\ref{empc:opt:3}).

        \item \parbox[t]{0.9\linewidth}{\eIf{$\Omega$ is empty}{ \vspace{1.5mm}
            Sort all the samples in ascending order with respect to the constraint costs $d(\hat{y}, \mathbb{Y})$. Let $\Lambda$ be the first $N_K$ elements. \vspace{1.5mm}
        }{ \vspace{1.5mm}
            {Sort the samples in $\Omega$ in ascending order with respect to the objective function values. Let $\Lambda$ be the first $N_K$ elements of $\Omega$ if the size of $\Omega$ is larger than $N_K$, otherwise let $\Lambda$ be $\Omega$.} \vspace{1.5mm}
        }}

        \item Obtain the parameters of the sampling distribution, $\mu_{k:k+N_c-1}$ and $\nu_{k:k+N_c-1}$, from the elite samples in $\Lambda$.

        \item Update the parameters of the sampling distribution for the next iteration \\  [0.07in]
                $\mu^{i+1}_{k:k+N_c-1} \leftarrow (1-\lambda) \mu_{k:k+N_c-1} + \lambda \mu^i_{k:k+N_c-1}$, \\  [0.07in]
                $\nu^{i+1}_{k:k+N_c-1} \leftarrow (1-\lambda) \nu_{k:k+N_c-1} + \lambda \nu^i_{k:k+N_c-1}$.

        \item  $i \leftarrow i+1$.

    \end{enumerate}
}}

\Return Optimal control action $u^*_k  \leftarrow u_{k|k}^{i-1,*}$, where $u_{k|k}^{i-1,*}$ is the first element of the sample sequence with lowest objective function values at the last iteration.
\end{algorithm*}

\textcolor{black}{A graphical illustration of the recursively updating process of the CE method is presented in Figure~\ref{fig:cem}, and the process of solving EMPC with the CE method~\cite{wen2018constrained,liu2020constrained} is described in Algorithm~\ref{CEM:MPC}. As shown in Figure~\ref{fig:cem}, to solve the optimization problem~(\ref{empc:opt}) using the CE method~\cite{wen2018constrained,liu2020constrained}, a control sequence~$u_{k|k}, \ldots, u_{k+N_c-1|k}$ that represents the potential solutions to~(\ref{empc:opt}) needs to be sampled. Given a maximum iteration~$N_{CE}$ and an initial sampling distribution~$\mathcal{N}\big(\mu^0_{k:k+N_c-1}, \nu^0_{k:k+N_c-1}\big)$, $N_{sample}$ random control sequences are sampled at each iteration, give as follows~\cite{wen2018constrained,liu2020constrained}:}
\begin{equation}\label{CEM:u}\small
    \Big\{\big(u_{k|k}^i, \ldots, u_{k+N_c-1|k}^i\big)_m\Big\}_{m=1}^{N_{sample}}  \sim \mathcal{N}\big(\mu^i_{k:k+N_c-1}, \nu^i_{k:k+N_c-1}\big),
\end{equation}
where $i = 0, \ldots, N_{CE}-1$; $\mathcal{N}\big(\mu^i_{k:k+N_c-1}, \nu^i_{k:k+N_c-1}\big)$ denotes $N_c$ separate multivariate Gaussian distributions, from which control input sequences are sampled; $\mu^i_{k:k+N_c-1}$ and $\nu^i_{k:k+N_c-1}$ denote the mean vector and covariance matrix, respectively. Given the sampled control sequences and current state $x_k$, the future states and outputs can be predicted according to~(\ref{empc:opt:1}) and~(\ref{empc:opt:3}), respectively. The objective function values, representing the accumulated economic costs, can then be computed based on these control sequences and their corresponding future states.

\textcolor{black}{As shown in Figure~\ref{fig:cem}(c), with the predicted future states and controlled outputs, the feasible set $\Omega$ of control sequences that satisfies the system constraints can be selected based on~(\ref{empc:opt:3}) and~(\ref{empc:opt:4}). Within the feasible set $\Omega$, we sort the samples in ascending order according to the accumulated cost, and we select the first $N_K$ samples (red dots shown in Figure~\ref{fig:cem}(d)) as an elite sample set $\Lambda$. If the feasible sample set $\Omega$ is empty, $N_K$ samples with the lowest constraint costs will be selected to construct the elite sample set $\Lambda$. The constraint cost is defined as the distance between the controlled outputs and the constraint set $\mathbb{Y}$, that is, $d(\hat{y}, \mathbb{Y}) = \text{inf}\big\{ \Vert \hat{y} - y \Vert_2 | y \in \mathbb{Y} \big\}$, where $\text{inf}$ is the infimum operator.}

% With the predicted future states and controlled outputs, the feasible set $\Omega$ of control sequences that satisfy the constraints can be selected. Within the feasible set $\Omega$, we sort the samples in an ascending order concerning the economic cost and select the first $N_K$ samples as an elite set $\Lambda$. If all the controlled outputs violate the output constraint, $N_K$ samples with the lowest constraint costs will be selected. The constraint cost is defined as the distance between the controlled output and the constraint set $\mathbb{Y}$, i.e., $d(\hat{y}, \mathbb{Y})=\inf\{\Vert \hat{y}-y\Vert_2~|~y \in \mathbb{Y}\}$ where $\inf$ is the infimum operator.

\textcolor{black}{The samples in the elite set are used to update the parameters of the sampling distribution for the next iteration. The parameters are updated by moving average as follows~\cite{wen2018constrained,liu2020constrained}:}
\begin{subequations}\label{CEM:update}
    \begin{align}
        \label{CEM:update:1} \mu^{i+1}_{k:k+N_c-1} &= (1-\lambda) \mu_{k:k+N_c-1} + \lambda \mu^i_{k:k+N_c-1}, \\
        \label{CEM:update:2} \nu^{i+1}_{k:k+N_c-1} &= (1-\lambda) \nu_{k:k+N_c-1} + \lambda \nu^i_{k:k+N_c-1},
    \end{align}
\end{subequations}
where $\mu_{k:k+N_c-1}$ and $\nu_{k:k+N_c-1}$ are the mean vector and covariance matrix of the control sequences in $\Lambda$; $\lambda$ is a parameter that adjusts the update of the sampling distribution. As $\lambda$ decreases, the updates converge faster to the new distribution.

\textcolor{black}{In the next iteration, the process is repeated using the updated distribution. The iteration continues until either the maximum number of iterations $N_{CE}$ is reached or the variance of the distribution becomes below a predefined minimum bound $v_{min}$. The first element $u^{i-1,*}_{k|k}$ of the optimal control sequence, which has the lowest accumulated cost at the last iteration, will be applied to the system. This way, the optimization problem~(\ref{empc:opt}) of EMPC at time instant $k$ is solved by the CE method~\cite{wen2018constrained,liu2020constrained}.}

% The iteration process will stop if it reaches the maximum iteration number $N_{CE}$ or the maximum variance drops below a minimum variance bound $\nu_{min}$. The first element $u_{k|k}^{i-1,*}$ of the optimal control sequence with the lowest economic cost at the last iteration will be applied to the system.

\section{Results}

In this section, we present the hybrid modeling and control results for the shipboard carbon capture process. Additionally, we conduct extensive comparisons to illustrate the superiority of the proposed approach.

\subsection{Settings}\label{scenario set}

\subsubsection{Parameter values}

\begin{table}[t!]
  \renewcommand\arraystretch{1.25}
  \caption{System parameters of shipboard carbon capture process.}\label{table:pcc:config}
  \centering\footnotesize
    \begin{tabular}{ l l l }
      \toprule
      \bfseries Parameter                            & \bfseries Value      \\ \midrule
      \bfseries Absorption and desorption column         \\
      Internal diameter of absorption column $D_c$  & 4.2~m \\
      Internal diameter of desorption column $D_c$  & 4.9~m \\
      Length of absorption column $l$               & 12.5~m \\
      Length of desorption column $l$               & 12.5~m \\
      Column interfacial area $a^I$   & 143.9~m$^2$/m$^3$ \\
      % Packing type                        & IMTP 40 \\
      Nominal packing size           & 0.038~m \\ \hline
      \bfseries Lean-rich solvent heat exchanger                  \\
      Volume of tube side $V_{tube}$   & 0.0155~m$^3$ \\
      Volume of shell side $V_{shell}$  & 0.4172~m$^3$ \\
      Overall heat transfer coefficient $U$  &1899.949~kW/K \\  \hline
      \bfseries Seawater heat exchanger       \\
      Inlet seawater temperature $T_{sw,in}$  &   308~K   \\
      Outlet seawater temperature $T_{sw,out}$  &     323~K   \\
      \textcolor{black}{Heat capacity of solvent $\tilde{C}_{p, sol}$} &  \textcolor{black}{3.9~kJ/(kg$\cdot$K)} \\
      \textcolor{black}{Heat capacity of seawater $\tilde{C}_{p, sw}$} &  \textcolor{black}{4.18~kJ/(kg$\cdot$K)} \\ \hline
      \bfseries Reboiler \\
      Holdup volume in reboiler $V_{reb}$    & 0.145~m$^3$ \\ \hline
      \bfseries Ship engines \\
      Engine output $Q_E$ with 100$\%$ engine load  & 10800~kW \\
      \textcolor{black}{Specific fuel oil consumption $W_{SFOC}$} & \textcolor{black}{0.1775~kg/kWh}\\  \hline
      \bfseries Fuel component \\
      Mole fraction of carbon in the fuel $q_{fuel,\text{C}}$ & $84.86\%$ \\ \hline
      \bfseries Flue gas component \\
      \textcolor{black}{Mole mass of carbon $r_{\text{C}}$} & \textcolor{black}{12.01~kg/kmol} \\
      \textcolor{black}{Mode mass of CO$_2$ $r_{\text{CO}_2}$} & \textcolor{black}{44.01~kg/kmol} \\
      Mole fraction of CO$_2$ in the flue gas $q_{flue,\text{CO}_2}$ & $5.462\%$ \\
      \bottomrule
    \end{tabular}
\end{table}

The system parameters of the shipboard post-combustion carbon capture process~(\ref{pcc:model}) are listed in Table~\ref{table:pcc:config}. The packing type is IMTP \#40. The parameters of the ship engines are adopted from~\cite{wartsila46f}. The fuel component is $n$-hexadecane~\cite{ben2023desulfurization}. The flue gas component in Table~\ref{table:pcc:config} is calculated from simulations conducted with Aspen Plus at different levels of ship engine load. The first-principles dynamic model introduced in Section~\ref{sec:system}, with the model parameters in Table~\ref{table:pcc:config}, serves as a high-fidelity process dynamic simulator, which can be used for data generation.

As discussed in Section~\ref{imFP}, the imperfect first-principles model contains inaccurate system parameters. Specifically, the constant coefficients for the gas and liquid phase mass transfer coefficients, derived from empirical formulations in~\cite{onda1968mass}, are set to~5.23 and~0.0051, respectively, in the simulator. However, in the imperfect first-principles model, these constants deviate to 3.08 and 0.0031. Furthermore, the imperfect first-principles model adjusts the interfacial heat transfer coefficient to 0.8 times its value in the simulator. Additionally, the enhancement factor for the desorption column in the imperfect first-principles model is 1.05 times the value adopted in the simulator.

The neural networks of the algebraic state inference and state dynamics compensation share a similar structure, each comprising three layers: an input layer, a hidden layer, and an output layer. Both neural networks utilize the hyperbolic tangent function as the activation function following the input and hidden layers. \textcolor{black}{For modeling performance comparison, two purely data-driven fully-connected feedforward neural networks~\cite{bebis1994feed} for the plant serve as the model baselines: one with 500 hidden neurons (NN1) and the other with 150 hidden neurons (NN2). The formulations of NN1 and NN2 are described as $\hat{\chi}^{\text{NN}1}_{k+1} = \text{NN}1 (x_k, z_k, u_k, p_k )$ and $\hat{\chi}^{\text{NN}2}_{k+1} = \text{NN}2 (x_k, z_k, u_k, p_k )$, respectively,
% \begin{equation*}\label{eq:nn}
%         {\chi}^{nn1}_{k+1} = F_{NN1} (x_k, z_k, u_k, p_k ), \quad
%         {\chi}^{nn2}_{k+1} = F_{NN2} (x_k, z_k, u_k, p_k )
% \end{equation*}
where $\hat{\chi}^{\text{NN}1}_{k+1} = [\hat{x}_{k+1}^{\text{NN}1 ^\top},  \hat{z}_{k+1}^{\text{NN}1 ^\top}]^\top$ denotes the output of the NN1 including the differential states and algebraic states at next time instant~$k+1$ (similarly for~$\hat{\chi}^{\text{NN}2}_{k+1}$). The two neural networks serve as dynamic models; they are used to predict both the algebraic states and the differential states for the next time instant. The inputs to both purely data-driven neural networks include differential states~$x_k$, algebraic states~$z_k$, control inputs~$u_k$, and the known disturbance~$p_k$ at current time instant~$k$.}
% Both neural networks take the differential and algebraic states, known disturbances, and control inputs as input and provide predictions for both the differential and algebraic states for the next time instant.
To ensure fair comparisons, the structures and most hyperparameters of the neural networks involved in the developed hybrid model remain the same as those of the two neural network-based model baselines, except for the number of neurons. The training batch size is set to 200, the training epoch is 1000, and the optimizer used is Adam with a learning rate of $10^{-4}$. The number of neurons in each layer of algebraic state inference and state dynamics compensation are 107-150-7 and 114-600-103, while those of NN1 and NN2 are 114-500-110 and 114-150-110, respectively.

\subsubsection{Simulation settings}

\begin{figure}[t!]
    \centering
    \includegraphics[width=0.45\textwidth]{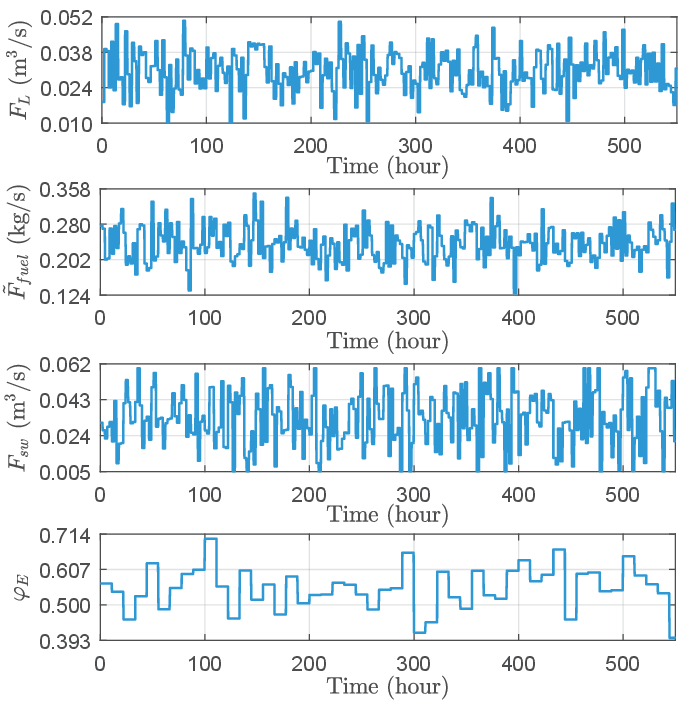}
   % \captionsetup{font={small}}
    \caption{Trajectories of the control inputs and the known disturbance generated for modeling.}\label{fig:result:up}
\end{figure}

\begin{table}[t!]
  \renewcommand\arraystretch{1.25}
  \caption{The upper bounds and lower bounds of control inputs $u$ and the known disturbance $p$.}\label{table:up:bound}
  \centering%\small
    \begin{tabular}{ c c c c c }
      \toprule
             &\parbox{.17\linewidth}{$F_L$ (m$^3$/s)} &\parbox{.19\linewidth}{$\tilde{F}_{fuel}$ (kg/s)} &\parbox{.18\linewidth}{$F_{sw}$ (m$^3$/s)} &\parbox{.08\linewidth}{$\varphi_E$} \\ \midrule
      \makecell{Upper \\bounds} &$0.04$        &$0.333$                    &$0.04$           &$1.0$\\
      \makecell{Lower \\bounds} &$0.02$        &$0.194$                     &$0.02$            &$0$\\ \bottomrule
    \end{tabular}
\end{table}

First, simulations of the process operation are conducted using the process dynamic simulator to generate data for hybrid modeling with a sampling period of $\Delta = 40$ s. The engine load varies after every 1000 sampling periods. The engine load level is determined following normal distribution with a mean of 0.55 and a standard deviation of 0.065, and is clipped in the bounded range. Open-loop control input sequences applied to the process are generated randomly from a uniform distribution within the prescribed ranges and are varied after every 200 sampling periods. The trajectories of the applied input sequences and the engine load sequence are shown in Figure~\ref{fig:result:up}. The upper bounds and lower bounds on the control inputs and engine load are shown in Table~\ref{table:up:bound}. The generated data is separated into training, validation, and test sets, which comprise 70\%, 10\%, and 20\% of the total data, respectively.

\subsubsection{Two evaluation cases}

In this study, we consider two cases for the evaluation of modeling performance. Case~I involves training data from three different ship operational conditions, where the engine load varies within different ranges. Table~\ref{table:engine load: exp} presents wide engine load ranges for different ship operational conditions considering the various ship types and sizes~\cite{faber2012regulated, pelic2023impact}. Conditions~1,~2, and~3 represent the slow steaming condition, the maneuvering condition, and the low engine load condition, respectively. Data collected under Condition~1 contributes 60\% of the training data, while Condition~2 and Condition~3 contribute 15\% and 25\%, respectively.
Case~II involves training data only from the slow steaming condition which is Condition~1, and the trained model will be tested on data from Condition~2 and Condition~3. Case~II is designed to analyze the advantages of the proposed hybrid modeling method in terms of predicting process states under unexplored operational conditions.

\begin{table}[t!]
    \renewcommand\arraystretch{1.2}
    \caption{Variability in engine load under different ship operational conditions.}\label{table:engine load: exp}
    \centering%\small
      \begin{tabular}{c c  }
        \toprule
        Operational condition     &Range of engine load  $\varphi_E$\\ \midrule
        Condition 1 (Slow steaming)  &40\% - 70\% \\
        Condition 2 (Maneuvering)  &80\% - 100\% \\
        Condition 3 (Low engine load)    &10\% - 30\% \\ \bottomrule
      \end{tabular}
  \end{table}

%\begin{rmk}
%The structures of the neural networks used for the algebraic state inference and state dynamics compensation can be different. In this study, feedforward neural networks have demonstrated excellent performance. While long short-term memory networks, designed to handle time dependencies, have been tested and shown similar results, their computational demands are considerably higher compared to feedforward neural networks. In addition, a single hidden layer proves sufficient to attain satisfactory modeling performance. Hence, we opt for a simple configuration to construct the algebraic state inference and state dynamics compensation for computational efficiency.
%\end{rmk}

\subsection{Modeling results for Case~I}

\subsubsection{Modeling performance}

\begin{figure}[t!]
    \centering
    \includegraphics[width=0.49\textwidth]{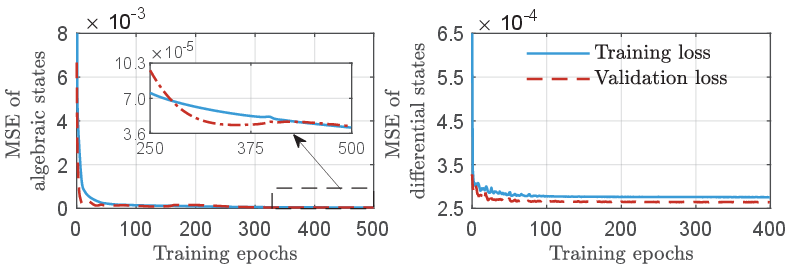}
   % \captionsetup{font={small}}
    \caption{Trajectories of the mean-squared error for algebraic state inference and state dynamics compensation based on the training and validation data sets.}\label{hybrid:fig:train loss:case1}
\end{figure}

\begin{figure}[t!]
    \centering
    \includegraphics[width=0.49\textwidth]{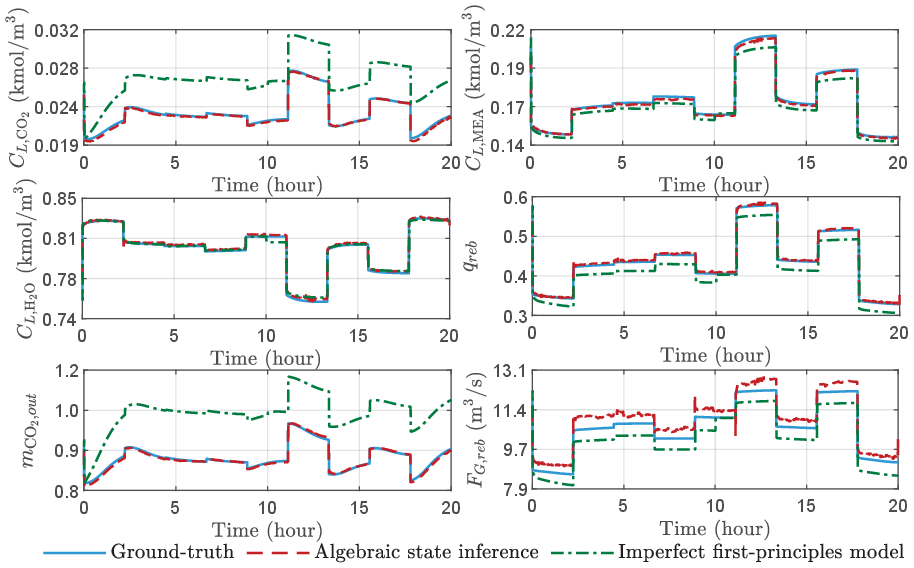}
   % \captionsetup{font={small}}
    \caption{Ground-truth and predictions of the algebraic states generated by algebraic state inference network and imperfect first-principles model under Case~I. The prediction horizon spans 1800 sample periods.}\label{fig:result:z:case1}
\end{figure}

\begin{figure}[t!]
    \centering
    \subfigure[Trajectories of ground-truth data and open-loop predictions of selected differential states from the absorption column under Case~I.]{
    \label{fig:result:abs:case1}
    \includegraphics[width=0.49\textwidth]{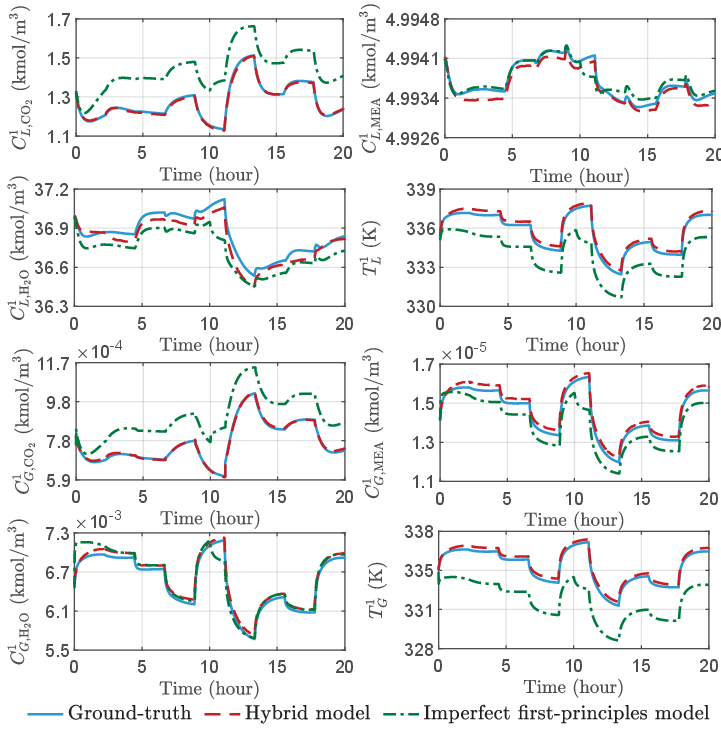}
    %\captionsetup{font={small}}
    }

    \subfigure[Trajectories of ground-truth data and open-loop predictions of selected differential states from the desorption column under Case~I.]{
    \label{fig:result:des:case1}
    \includegraphics[width=0.49\textwidth]{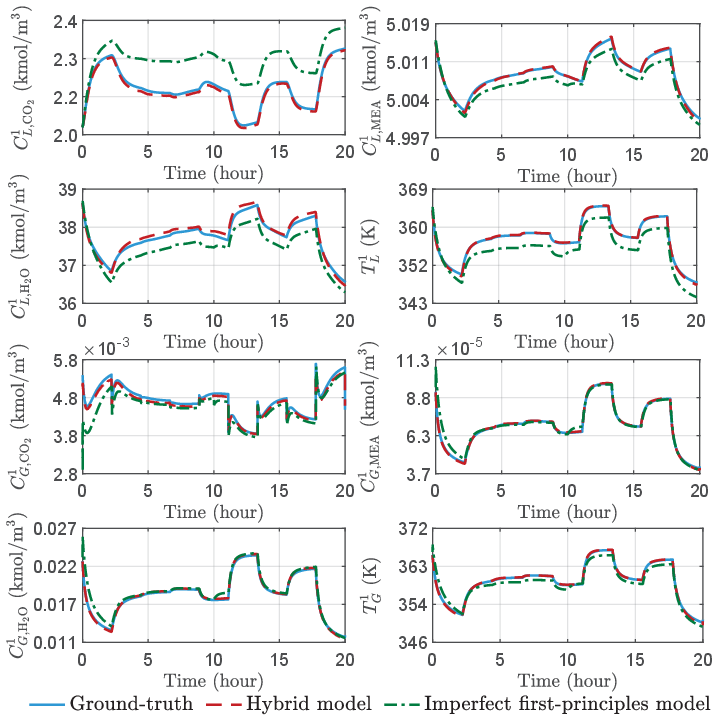}
    }
    \caption{Trajectories of ground-truth data and open-loop predictions of selected differential states from the absorption and desorption columns under Case~I. The prediction horizon spans 1800 sampling periods.}
    \label{fig:result:x:case1}
\end{figure}

Within the hybrid model framework, the algebraic state inference and state dynamics compensation neural networks are trained separately. The trajectories of the mean-squared error (MSE) for prediction on both training and validation data sets, encompassing algebraic and differential states in each epoch, are presented in Figure~\ref{hybrid:fig:train loss:case1}. The MSE is defined as $\text{MSE}=\frac{1}{N_{dim}N_{data}} \sum_{i=1}^{N_{data}}\big\Vert r_i - \hat{r}_i \big \Vert_2^2$, where $r$ and $\hat{r}$ are the real state and predicted state, respectively; $N_{dim}$ is the dimensions of the state; $N_{data}$ is the number of data sets. We compute MSEs for the training, validation, and test data sets, and $N_{data}$ takes the value of $3.5\times10^4$, $5\times10^3$, and $10^4$, respectively. Considering the significantly different magnitudes of the process states, the MSEs are calculated using normalized system state values. As shown in Figure~\ref{hybrid:fig:train loss:case1}, both the neural networks for algebraic state inference and state dynamics compensation converge as training progresses.

\textcolor{black}{Figure~\ref{fig:result:z:case1} presents the ground-truth data obtained from the simulator, and the prediction results of the algebraic states obtained by the algebraic state inference network and the imperfect first-principles model. The prediction horizon spans 1800 sample periods, which is equivalent to 20 hours. The prediction results of the algebraic state inference (shown in red dashed line) can accurately track the ground-truth generated by the simulator (shown in blue line), and have much better accuracy in comparison with the predictions of the imperfect first-principles model (shown in green dashed dotted line). The discrepancy between the results of the imperfect first-principles model and the simulator arises from inaccuracy in the parameters.}

% The predictions of the algebraic state inference, shown in the red dashed lines, can accurately track the trends of actual state trajectories. For some state variables including $C_{L, \text{CO}_2}$ and $m_{\text{CO}_2, out}$ in Figure~\ref{fig:result:z:case1}, the discrepancies between the predictions of the imperfect first-principles model and the ground-truth are significantly larger than those produced by the established hybrid model.

\begin{table}[!t]
    \renewcommand\arraystretch{1.2}
    \caption{MSE results for algebraic and differential state predictions of the proposed hybrid model and imperfect first-principles model compared with ground-truth generated by simulator under Case~I.}\label{table:case1:model:mse}
    \centering%\small
      \begin{tabular}{c c c c c}
        \toprule
           & Hybrid model  &\makecell{Imperfect first-\\principles model} \\ \midrule
        Algebraic states $z$   &0.0035 &0.0084  \\
        Differential states $x$   &0.0043  &0.0528 \\  \bottomrule
      \end{tabular}
  \end{table}

Figure~\ref{fig:result:x:case1} presents ground-truth data and open-loop trajectories of selected differential states from the absorption and desorption columns. These state variables include the molar concentrations in the liquid and gas phases of CO$_2$, MEA, and H$_2$O, along with the temperatures in the liquid and gas phases. \textcolor{black}{As shown in Figure~\ref{fig:result:x:case1}, the hybrid model can track the ground-truth generated by the simulator accurately and provide higher prediction accuracy than the imperfect first-principles model. The MSE results for the proposed hybrid model and imperfect first-principles model are presented in Table~\ref{table:case1:model:mse}. The proposed hybrid model reduces the prediction errors for algebraic states and differential states by 58.33\% and 91.86\%, respectively, compared to the imperfect first-principles model.}

\subsubsection{Comparisons}\label{section:datasize:case1}

In this subsection, comparisons are conducted between the hybrid model built based on the proposed method and the two neural networks (NN1 and NN2) built using fully-connected feedforward neural network~\cite{bebis1994feed} to analyze the modeling performance regarding data efficiency and model robustness.

To examine the modeling performance of the three models in terms of data efficiency, four training data sets of different sizes are explored: 5$\times 10^4$, $10^5$, 1.5$\times 10^5$, and 2$\times 10^5$ samples. With each data set, the hybrid model, NN1 model, and NN2 model are trained. All the models are validated using the same test data set, and the corresponding MSEs are computed for comparative analysis.

\begin{figure}[t!]
    \centering
    \includegraphics[width=0.48\textwidth]{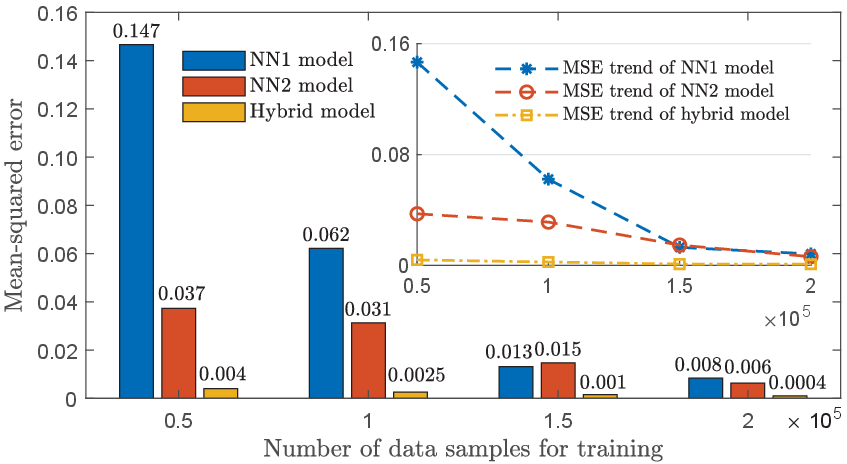}
   % \captionsetup{font={small}}
    \caption{MSEs of 1800-step-ahead open-loop predictions of different models built under Case~I, calculated using the test data set.}\label{fig:result:error:datasize:case1}
\end{figure}

Figure~\ref{fig:result:error:datasize:case1} compares the MSEs for the models based on 1800-step-ahead open-loop predictions of the differential states generated by the models subject to different training data sizes, and presents the trends of MSEs with data increase. As shown in Figure~\ref{fig:result:error:datasize:case1}, the MSEs of the three models decrease with the increase in the data size, which aligns with intuitive expectations. The proposed hybrid model demonstrates significantly smaller prediction error compared to the baseline neural network models.
Moreover, with considerably smaller training data sizes, the hybrid model demonstrates comparable performance to that of neural network models trained with larger data sets, which showcases the superiority of the proposed method in terms of data efficiency.

To evaluate the robustness of the proposed hybrid model, individual tests are conducted under the three conditions. Models trained with an equal amount of data are chosen for comparison: the hybrid model and the NN2 model with $2\times 10^5$ training data samples.
Table~\ref{table:case1:mse} presents the MSEs of the hybrid model and the NN2 model under the three conditions.
The hybrid model demonstrates consistent modeling performance across various operational conditions. The NN2 model exhibits significantly poorer performance under Condition 3, particularly when the engine load is in a slow range. This indicates that the proposed hybrid model offers enhanced robustness, benefiting from the integration of first-principles knowledge.

\begin{table}[t!]
  \renewcommand\arraystretch{1.2}
  \caption{MSEs of predictions for the hybrid model and NN2 model under different ship operational conditions, calculated using the test data set.}\label{table:case1:mse}
  \centering%\small
    \begin{tabular}{c  c c c  }
      \toprule
          &Condition 1 &Condition 2 &Condition 3 \\ \midrule
      Hybrid model    &0.0004 &0.0020  & 0.0004 \\
      NN2 model    &0.0063 &0.0073 & 0.0918 \\  \bottomrule
    \end{tabular}
\end{table}

\begin{figure}[t!]
    \centering
    \includegraphics[width=0.49\textwidth]{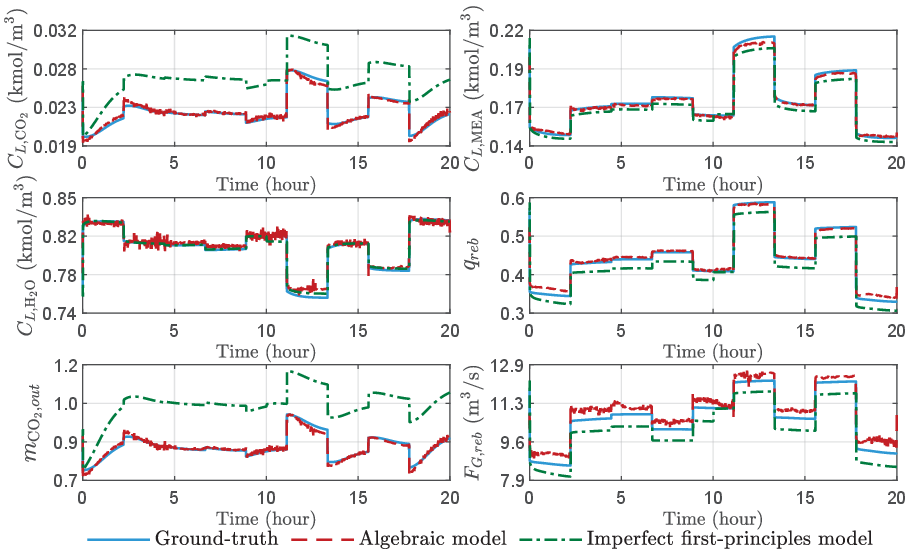}
   % \captionsetup{font={small}}
    \caption{Ground-truth and predictions of the algebraic states generated by algebraic state inference network and imperfect first-principles model under Case~II. The prediction horizon spans 1800 sample periods.}\label{fig:result:z:case2}
\end{figure}

\begin{figure}[t!]
    \centering
    \subfigure[Trajectories of ground-truth data and open-loop predictions of selected differential states from the absorption column under Case~II.]{
    \label{fig:result:abs}
    \includegraphics[width=0.49\textwidth]{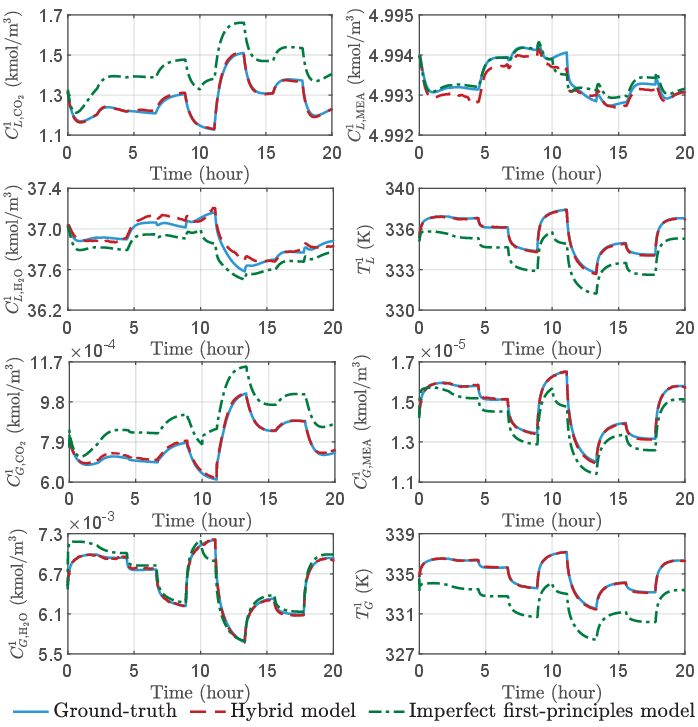}
    %\captionsetup{font={small}}
    }

    \subfigure[Trajectories of ground-truth data and open-loop predictions of selected differential states from the desorption column under Case~II.]{
    \label{fig:result:des}
    \includegraphics[width=0.49\textwidth]{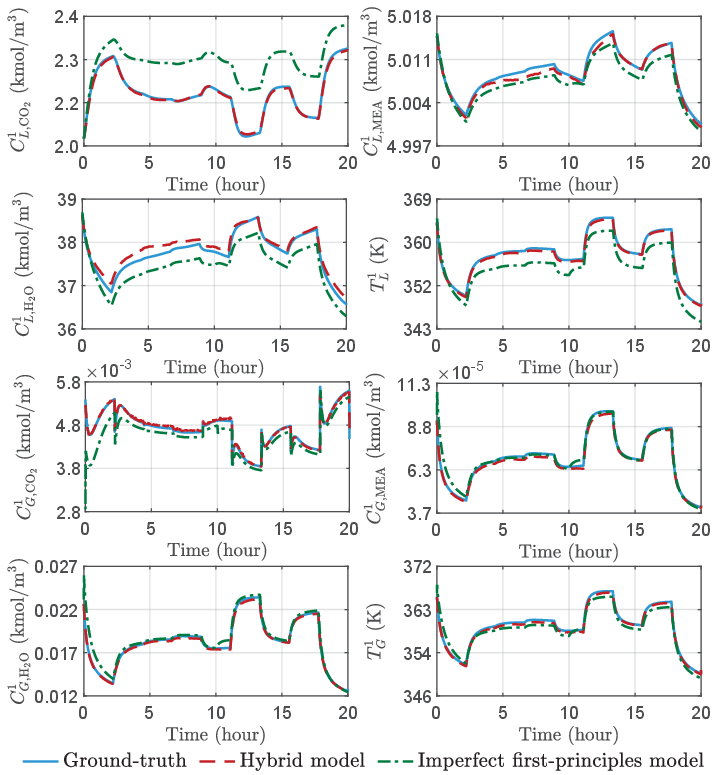}
    }
    \caption{Trajectories of ground-truth data and open-loop predictions of selected differential states from the absorption and desorption columns under Case~II. The prediction horizon spans 1800 sampling periods.}
    \label{fig:result:x}
\end{figure}

\subsection{Modeling results for Case~II}

In Case~II, we construct a hybrid model by using only the data collected from Condition 1 (slow steaming condition), and evaluate its performance.

\subsubsection{Modeling performance}

\begin{figure}[t!]
    \centering
    \includegraphics[width=0.48\textwidth]{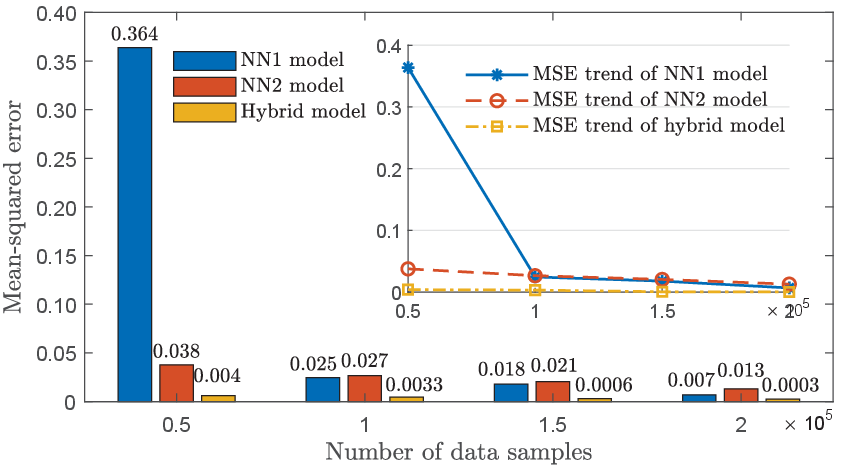}
   % \captionsetup{font={small}}
    \caption{MSEs of 1800-step-ahead open-loop predictions of the hybrid model and NN1 and NN2 models~\cite{bebis1994feed} built under Case~II, calculated using the test data set.}\label{fig:result:error:datasize}
\end{figure}

\begin{figure*}[t!]
    \centering
    \includegraphics[width=0.83\textwidth]{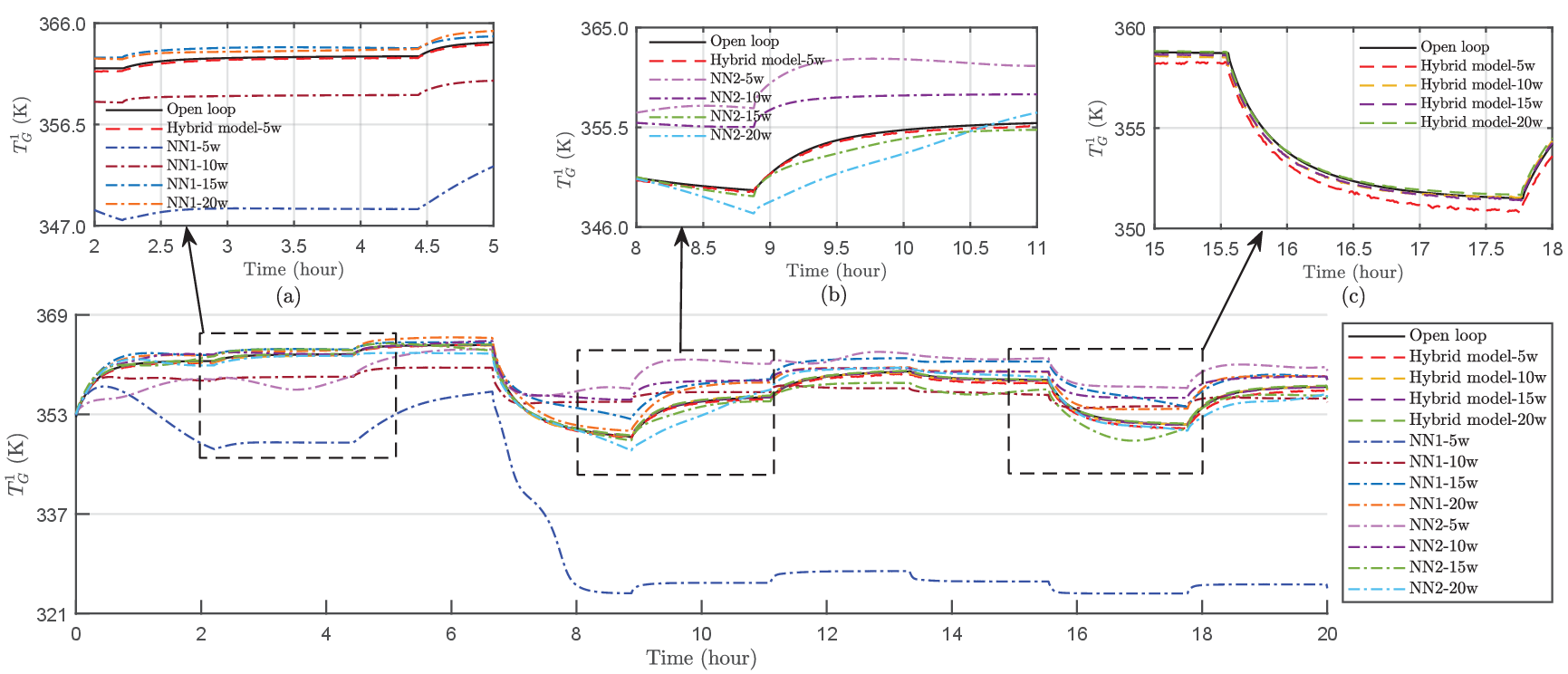}
   % \captionsetup{font={small}}
    \caption{Trajectories of open-loop predictions for gas temperature in the desorption column generated by the hybrid models and neural network models under different training data sizes. Subplots (a) and (b) present the prediction trajectories of the NN1 and NN2 models based on fully-connected feedforward neural network~\cite{bebis1994feed} under different training data sizes, respectively. The state trajectory of the hybrid model trained with 5$\times 10^4$ is added for comparison. Subplot (c) presents the prediction trajectories of the hybrid models under different training data sizes.}\label{fig:result:state:datasize}
\end{figure*}

\begin{figure}[t!]
    \centering
    \subfigure[Ground-truth and predictions generated by the hybrid model and NN1 model for Condition 2.]{
    \label{fig:result:state:generalzation:p810}
    \includegraphics[width=0.42\textwidth]{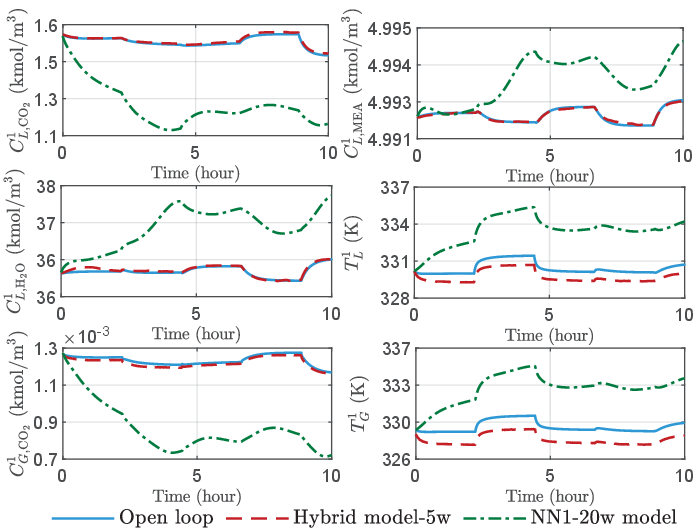}
    %\captionsetup{font={small}}
    }

    \subfigure[Ground-truth and predictions generated by the hybrid model and NN1 model for Condition 3.]{
    \label{fig:result:state:generalzation:p13}
    \includegraphics[width=0.42\textwidth]{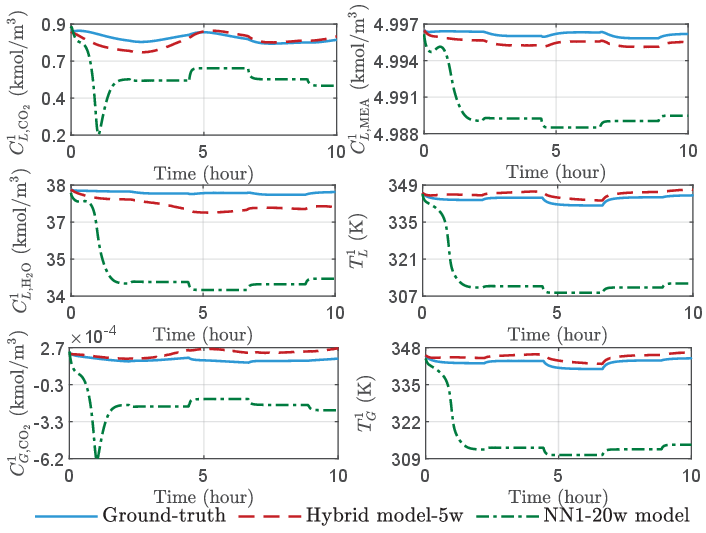}
    }
    \caption{Trajectories of ground-truth data and open-loop predictions for two unexplored operational conditions.}
    \label{fig:result:state:generalzation}
\end{figure}

\textcolor{black}{Figure~\ref{fig:result:z:case2} presents the ground-truth data obtained by the simulator, and the prediction results of the algebraic states given by the algebraic state inference network and the imperfect first-principles model.} The prediction horizon spans 1800 sample periods. The predictions of the algebraic state inference, shown in red dashed lines, can accurately track the trends of actual state trajectories. \textcolor{black}{The proposed hybrid model demonstrates notably accurate tracking performance, especially for algebraic states $C_{L, \text{CO}_2}$ and $m_{\text{CO}_2, out}$, compared to the imperfect first-principles model.}
% The predictions of $C_{L, \text{CO}2}$ and $m_{\text{CO}_2, \text{out}}$ generated by the imperfect first-principles model exhibit notably inferior tracking performance compared to the proposed hybrid model.

Figure~\ref{fig:result:x} presents ground-truth data and open-loop prediction trajectories of selected differential states from the absorption and desorption columns by different models. The hybrid model can accurately track the actual differential state trajectories. As shown in Figure~\ref{fig:result:x}, the predictions of the hybrid model are significantly more accurate than the predictions of the imperfect first-principles model.

\subsubsection{Data efficiency}\label{section:datasize}

We consider various training data sizes obtained from Condition~1 (the slow steaming condition), and for each data size, we establish a hybrid model, an NN1 model, and an NN2 model. The sizes of training data sets are the same as those for Case~I.

Figure~\ref{fig:result:error:datasize} presents the MSEs for the three models based on 1800-step predictions of differential states of different models, and displays the trends of MSEs with data increase. As shown in Figure~\ref{fig:result:error:datasize}, the proposed hybrid model demonstrates notably reduced prediction errors, showcasing superior data efficiency when compared to the purely data-driven NN1 and NN2 models built based on fully-connected feedforward neural network~\cite{bebis1994feed}, which does not contain first-principles knowledge. Moreover, all three models exhibit decreasing MSEs as the training data size increases.

% To evaluate the modeling performance regarding different models and different training data sizes, Figure~\ref{fig:result:state:datasize} provides the open-loop prediction trajectories of gas temperature in the desorption column. Three subplots, labeled as (a), (b), (c) in Figure~\ref{fig:result:state:datasize}, present the comparison of the modeling performance among the NN1 model, NN2 model, and hybrid model across different training data sizes, respectively. Specifically, Figure~\ref{fig:result:state:datasize}(a) and~\ref{fig:result:state:datasize}(b) illustrate the NN model can not achieve accurate multi-step tracking performance with limited training data. As shown in Figure~\ref{fig:result:state:datasize}(c), the hybrid model demonstrates satisfactory modeling performance with small training data sizes.

\textcolor{black}{To further compare the prediction performance across different modeling approaches and training data sizes, Figure~\ref{fig:result:state:datasize} presents the ground-truth data and open-loop prediction trajectoris for the gas temperature in the desorption column provided by different models. This results in a total of 15 distinct models using various modeling methods and training data sets. To facilitate easy expression, we use suffixes 5w, 10w, 15w, and 20w to represent the training data size of the model, corresponding to $5 \times 10^4$, $10^5$, $1.5 \times 10^5$, and $2 \times 10^5$ samples, respectively. For example, ``Hybrid model-10w'' represents the hybrid model built based on the proposed method with $10^5$ data samples.} The three subplots, labeled as (a), (b), (c) in Figure~\ref{fig:result:state:datasize}, present the comparison of the modeling performance among the NN1 model, NN2 model, and hybrid model across different training data sizes, respectively. Specifically, from Figures~\ref{fig:result:state:datasize}(a) and~\ref{fig:result:state:datasize}(b), the purely data-driven NN1 and NN2 models based on fully-connected feedforward neural network~\cite{bebis1994feed} cannot achieve accurate multi-step tracking performance with limited training data. In contrast, Figure~\ref{fig:result:state:datasize}(c) demonstrates that the hybrid model maintains satisfactory modeling performance even with small training data sizes.

\subsubsection{Generalizability}

To evaluate the modeling performance under unexplored operational conditions, we test the hybrid model and neural network models that are trained using data from Condition 1 on Condition 2 and Condition 3. To ensure fair comparison, we select models with comparable performance under Condition 1: the hybrid model trained with $5\times10^4$ data samples and the NN1 model trained with $2\times10^5$ data samples.

Figure~\ref{fig:result:state:generalzation} presents the ground-truth data and predictions obtained by the hybrid model and NN1 model for two unexplored operational conditions. For Condition 2, as shown in Figure~\ref{fig:result:state:generalzation:p810}, some of the state predictions of the hybrid model, shown in red dashed lines, can accurately track the actual trajectories, while others have noticeable deviations. The predictions of the NN1 model fail to track the actual trajectories. For Condition 3, as shown in Figure~\ref{fig:result:state:generalzation:p13}, the predictions of the hybrid model have a relatively smaller deviation from the actual trajectories in comparison with those of the NN1 model.
%{\color{red}It is worth noting that in Figure~\ref{fig:result:state:generalzation:p13}, some prediction values of the state variables are negative. This occurs because the output layer of the neural network does not use an activation function. }

\begin{table}[t!]
  \renewcommand\arraystretch{1.2}
  \caption{MSEs of predictions for the hybrid model and NN1 model under different ship operational conditions, calculated using the test data set.}\label{table:case2:mse}
  \centering%\small
    \begin{tabular}{c c c c }
      \toprule
           &Condition 1 &Condition 2 &Condition 3 \\ \midrule
      Hybrid model    &0.0040 &0.0101  & 0.1852 \\
      NN1 model    &0.0068 &0.0313 & 0.8784 \\  \bottomrule
    \end{tabular}
\end{table}

\begin{figure}[t!]
    \centering
    \includegraphics[width=0.48\textwidth]{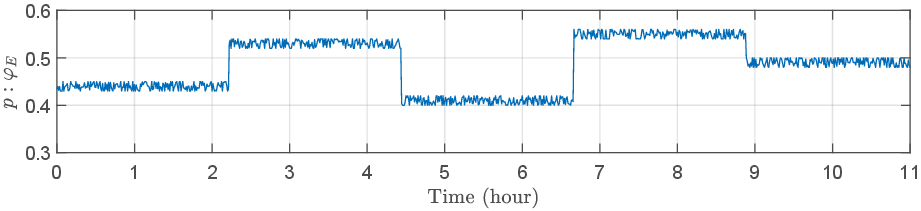}
   % \captionsetup{font={small}}
    \caption{Trajectory of the engine load.}\label{fig:control:p}
\end{figure}

Table~\ref{table:case2:mse} displays the MSEs of both the selected hybrid model and NN1 model across different conditions. The hybrid model consistently exhibits significantly lower prediction errors across all three conditions compared to the NN1 model. This suggests that the proposed hybrid model possesses superior generalization capability.

\subsection{Control results}\label{sec:control results}

In the design of the hybrid model-based economic model predictive control, the carbon tax is $\alpha = \$0.05$/kg~\cite{decardi2018improving}, the fuel price is $\beta= \$1.2852$/kg~\cite{luo2017study}, the limit value for CO$_2$ emissions is $y_{limit} = 0.5$ kg/s, and the control horizon is $N_c = 5$. \textcolor{black}{The control inputs generated by the controller are updated every 10 sampling periods (400 seconds) to ensure sufficient time for solving the optimization problem of the EMPC.} The control inputs are subject to constraint with the ranges listed in Table~\ref{table:up:bound}. An output constraint is imposed on the reboiler temperature to ensure solvent performance, with the reboiler temperature confined to 385.15~K $\leq y(2) \leq 393.15$~K. In terms of the execution of the CE method, the maximum iteration number is $N_{CE}=20$, the sample numbers generated each iteration is $N_{sample}=400$, the size of the elite samples is $N_K=20$, the update rate $\lambda =0.01$, and the minimum variance bound is $\nu_{min}=10^{-8}$. The parameters of the initial sampling distribution are $\mu^0_{k:k+N_c-1}=[0.03, 0.2635, 0.03]^{\top}$ and $\nu^0_{k:k+N_c-1}=I_{3\times3}$, where $I$ is an identity matrix. The mean vector values correspond to the medians of the bounded range for the control inputs. Figure~\ref{fig:control:p} displays the trajectory of the time-varying engine load, which is considered as the known disturbance to the process.

\begin{figure}[!t]
    \centering
    \includegraphics[width=0.47\textwidth]{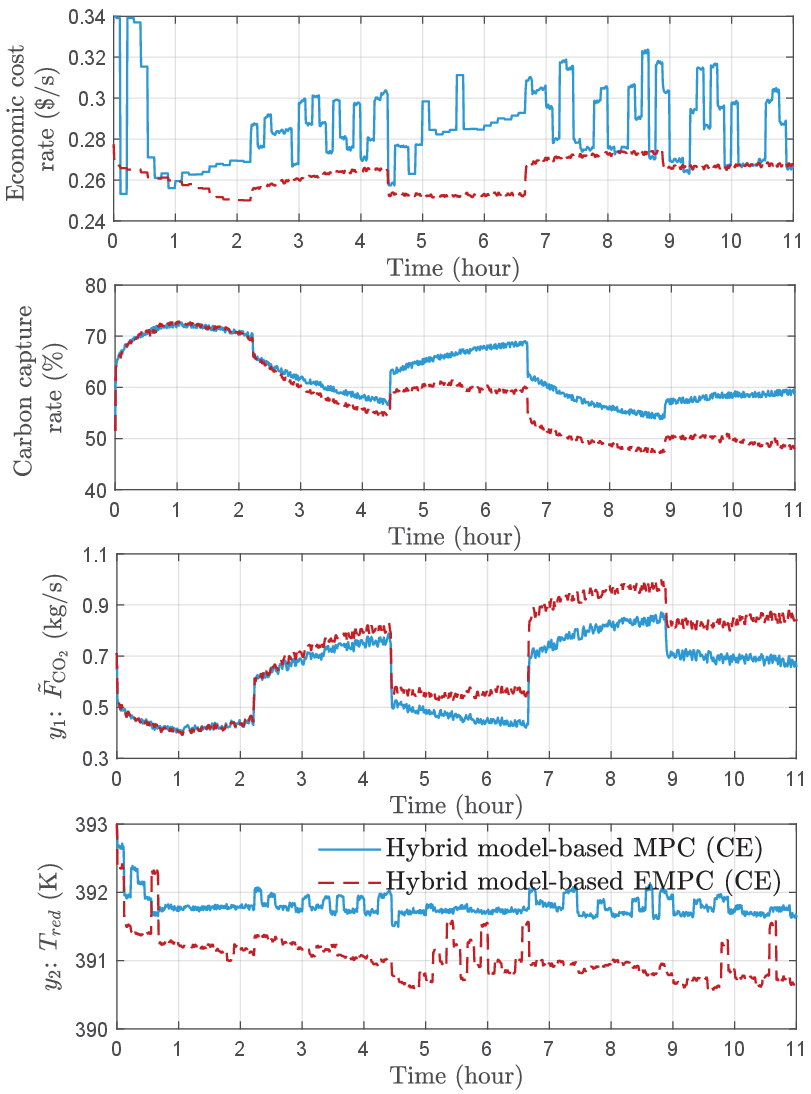}
    \caption{Closed-loop trajectories of economic cost rate, carbon capture rate, and controlled outputs with hybrid model-based MPC and EMPC designs implemented with the CE method~\cite{wen2018constrained,liu2020constrained}.}\label{fig:control:controller:ey}
\end{figure}

\begin{figure}[!t]
    \centering
    \includegraphics[width=0.47\textwidth]{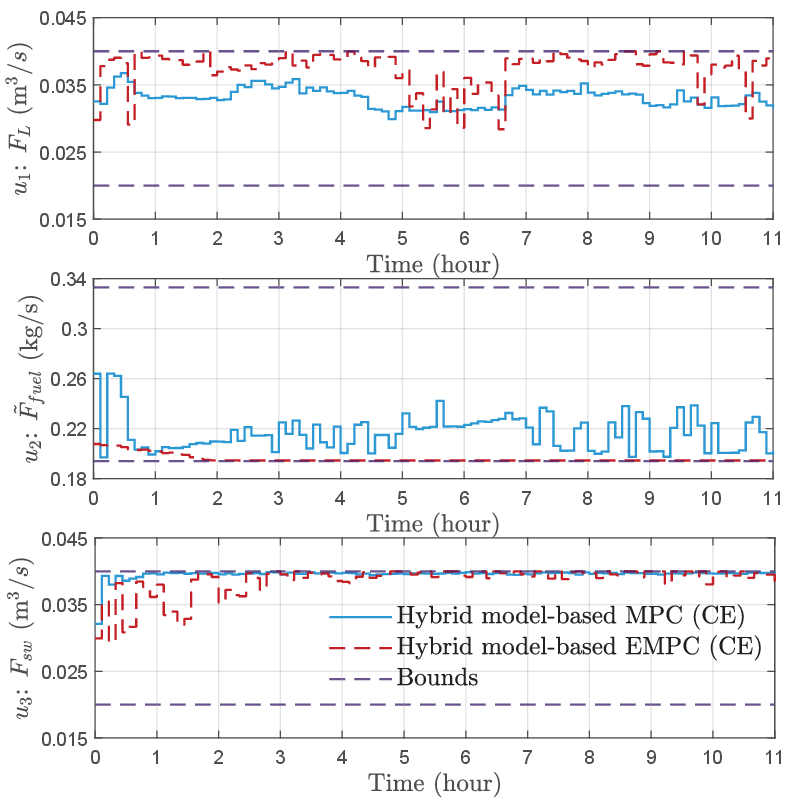}
    \caption{Closed-loop trajectories of control inputs obtained from hybrid model-based MPC and EMPC designs.}\label{fig:control:controller:u}
\end{figure}

\subsubsection{Comparison with different control schemes}

\textcolor{black}{To evaluate the performance of the proposed hybrid model-based EMPC design, a conventional set-point tracking MPC is considered for comparison. The objective function of MPC is a quadratic cost function that penalizes the deviations of the states from desired set-points, that is, $\ell_{mpc}(y_k, u_k) = \big\Vert y_s - \hat{y}_{k}\big\Vert_{Q_{mpc}}^2 + \big \Vert u_s - u_{k} \big\Vert_{R_{mpc}}^2 $, where $y_s$ is the set-point for the system outputs; $u_s$ is the steady-state reference input corresponding to $y_s$; $Q_{mpc}$ and $R_{mpc}$ are the weighting matrices for the outputs and control inputs. The set-point is determined through steady-state optimization, with the objective function being the economic cost in~(\ref{empc:cost}).} The weighting matrices of the output variables and control inputs are set as $Q_{mpc}=\text{diag}([3, 10]) \in \mathbb{R}^{2\times 2}$ and $R_{mpc}=\text{diag}([0.08, 0.08, 0.08])\in \mathbb{R}^{3\times 3}$, respectively, where $\text{diag}(\cdot)$ denotes a diagonal matrix.

% \begin{figure*}[t!]
%     \centering
%     \includegraphics[width=1\textwidth]{Fig/241_ux_2.pdf}
%    % \captionsetup{font={small}}
%     \caption{Closed-loop trajectories of controlled output, economic cost, and control inputs with MPC and EMPC designs.}\label{fig:control:241}
% \end{figure*}

\textcolor{black}{Figure~\ref{fig:control:controller:ey} presents the closed-loop trajectories of the economic cost rate, carbon capture rate, and controlled outputs under hybrid model-based MPC and EMPC designs implemented with the CE method~\cite{wen2018constrained,liu2020constrained}. Figure~\ref{fig:control:controller:u} presents the corresponding control input trajectories along with input constraints. The average carbon capture rates based on the MPC and EMPC designs are 62.69\% and 57.77\%, respectively. The lower carbon capture rate of the EMPC design is attributed to its lower fuel flow rate $\tilde{F}_{fuel}$ for the diesel gas turbine, as shown in Figure~\ref{fig:control:controller:u}. The differences in fuel flow rate and liquid solvent flow rate lead to a lower reboiler temperature $T_{reb}$, as shown in Figure~\ref{fig:control:controller:ey}. Consequently, there is less CO$_2$ released from the recycled solvent in the reboiler during the thermal regeneration process. This decrease results in a lower amount of CO$_2$ absorbed by the recycled solvent in the absorption column, which leads to a reduced carbon capture rate.}
%Figure~\ref{fig:control:241} presents the closed-loop trajectories of the controlled outputs and the economic cost, as well as the trajectories of the optimal control inputs obtained based on the MPC and EMPC schemes. As shown in Figure~\ref{fig:control:241}, the MPC design is capable of regulating the state to the set-point. Meanwhile, the proposed EMPC scheme can find a relatively more cost-saving operation without violating the constraints. The average economic cost rate of the shipboard carbon capture process with MPC and EMPC designs are \$0.285/s and \$0.262/s, respectively. These results demonstrate an 8.07\% reduction in the overall operational economic cost with the utilization of the proposed EMPC, as compared to the commonly adopted set-point tracking MPC.

\textcolor{black}{Meanwhile, we note that the proposed EMPC design (shown in the red dashed lines in Figure~\ref{fig:control:controller:ey}) enables a more cost-saving operation when operating at an average carbon capture rate lower than that of the MPC design. The average economic cost rates of the shipboard PCC system are \$0.285/s when using the MPC design and \$0.262/s when using the EMPC design; these results indicate an 8.07\% reduction in overall operational costs achieved by the proposed EMPC approach, as compared to the commonly adopted set-point tracking MPC. This is because the objective function of the EMPC design considers the dynamic operational cost of the shipboard PCC plant, including not only the tax for emitted CO$_2$ but also the fuel cost. Additionally, the tax for the emitted CO$_2$ is relatively small compared to the fuel cost in the economic cost function.}

\subsubsection{Comparison with different models}

\textcolor{black}{Additionally, we have compared the proposed hybrid model-based EMPC with the imperfect first-principles model-based EMPC. Both the proposed hybrid model-based EMPC design and the imperfect first-principles model-based EMPC are solved using the CE method~\cite{wen2018constrained,liu2020constrained}, with the same settings as mentioned at the beginning of Subsection~\ref{sec:control results}.}

\begin{figure}[!t]
    \centering
    \includegraphics[width=0.47\textwidth]{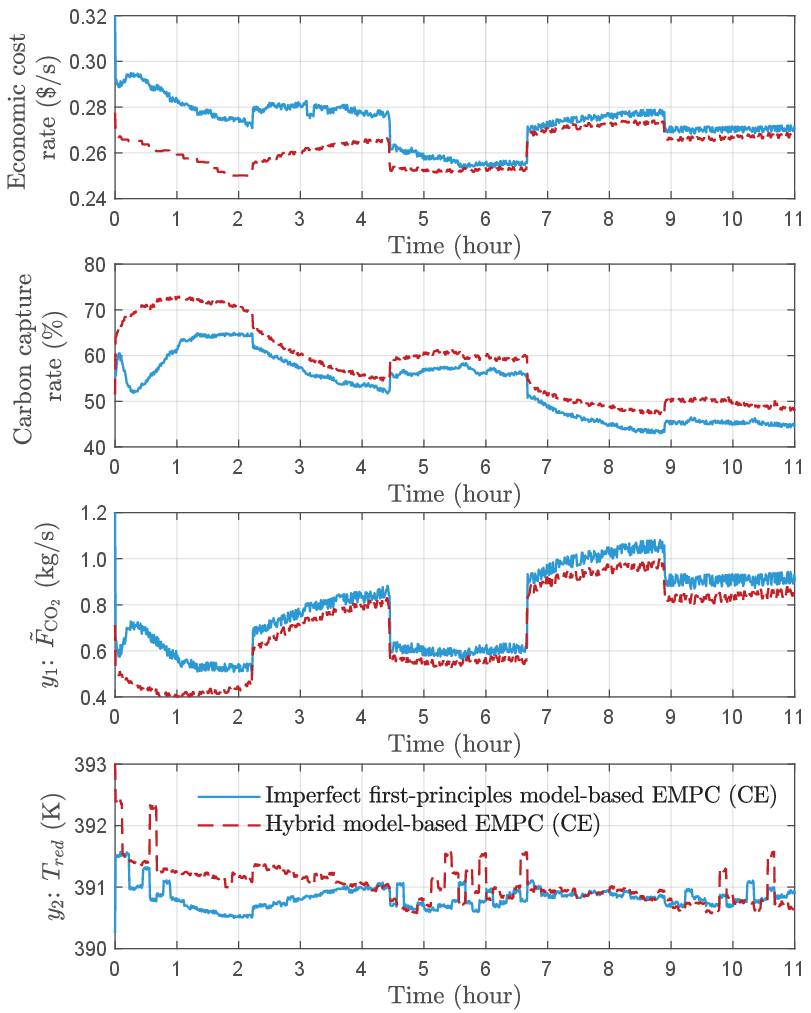}
    \caption{Closed-loop trajectories of economic cost rate, carbon capture rate, and controlled outputs with hybrid model-based EMPC and imperfect first-principles model-based EMPC designs implemented with the CE method~\cite{wen2018constrained,liu2020constrained}.}\label{fig:control:model:ey}
\end{figure}

\begin{figure}[!t]
    \centering
    \includegraphics[width=0.47\textwidth]{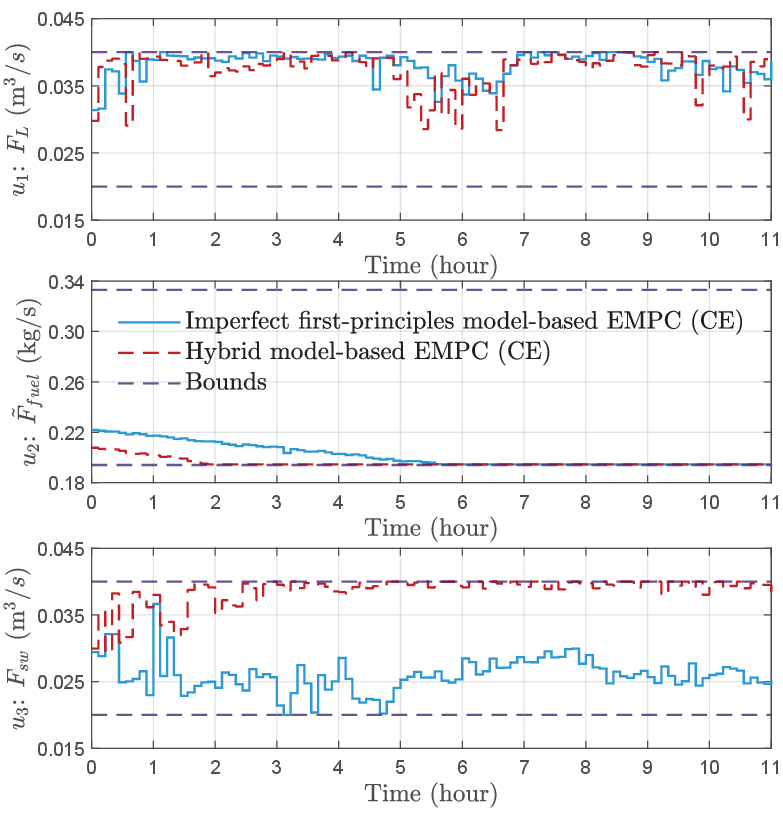}
    \caption{Closed-loop trajectories of control inputs obtained from hybrid model-based EMPC and imperfect first-principles model-based EMPC designs.}\label{fig:control:model:u}
\end{figure}

\begin{table}[!t]
    \renewcommand\arraystretch{1.2}
    \caption{Comparison results of the EMPC approaches based on the proposed hybrid model and the imperfect first-principles model.}\label{table:compare:model}
    \centering\footnotesize%\small
      \begin{tabular}{c c c}
        \toprule
             &  \makecell{Hybrid model-\\based EMPC} & \makecell{Imperfect first-principles \\ model-based EMPC} \\ \midrule
         \makecell{Economic cost\\rate (\$/s)}  &0.262  &0.273   \\
         \makecell{Carbon capture\\rate (\%)}  &57.77  &52.95  \\  \bottomrule
      \end{tabular}
  \end{table}

\textcolor{black}{Figure~\ref{fig:control:model:ey} presents the closed-loop trajectories of the economic cost rate, the carbon capture rate, and the controlled outputs obtained by the proposed hybrid model-based EMPC and the imperfect first-principles model-based EMPC. Figure~\ref{fig:control:model:u} presents their corresponding control input trajectories given by the two designs. As shown in Figure~\ref{fig:control:model:ey}, the economic cost rate of the EMPC design based on the imperfect first-principles model is significantly higher compared to that of the proposed hybrid model-based EMPC. Additionally, the carbon capture rate obtained based on the proposed hybrid model-based EMPC is higher than that of the imperfect first-principles model-based EMPC. Specifically, the average economic cost rate and carbon capture rate of these two designs are listed in Table~\ref{table:compare:model}. Based on the results, the proposed hybrid model EMPC acheives a 4.20\% reduction in overall operational economic cost and a 9.10\% improvement in carbon capture rate compared to the imperfect first-principles model-based EMPC.}

\subsubsection{Comparison with different optimization solvers}

\textcolor{black}{We also compare the control performance and computation times of the proposed EMPC method implemented using the CE method~\cite{wen2018constrained,liu2020constrained}, the interior point optimizer (IPOPT) solver~\cite{wachter2006implementation}, and the sequential quadratic programming (SQP) solver~\cite{matlab}.}

\begin{figure}[!t]
    \centering
    \includegraphics[width=0.47\textwidth]{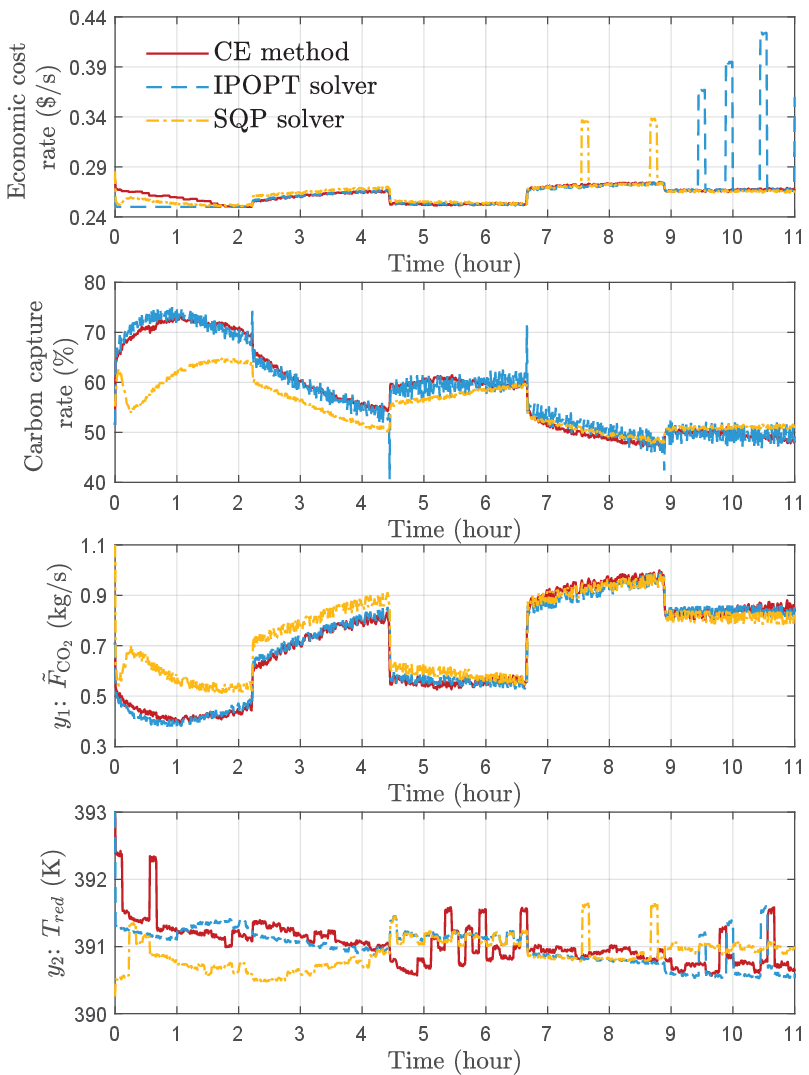}
    \caption{Closed-loop trajectories of economic cost rate, carbon capture rate, and controlled outputs with hybrid model-based EMPC solved by the CE method~\cite{wen2018constrained,liu2020constrained}, the IPOPT solver~\cite{wachter2006implementation}, and the SQP solver~\cite{matlab}.}\label{fig:control:ey}
\end{figure}

\begin{figure}[!t]
    \centering
    \includegraphics[width=0.47\textwidth]{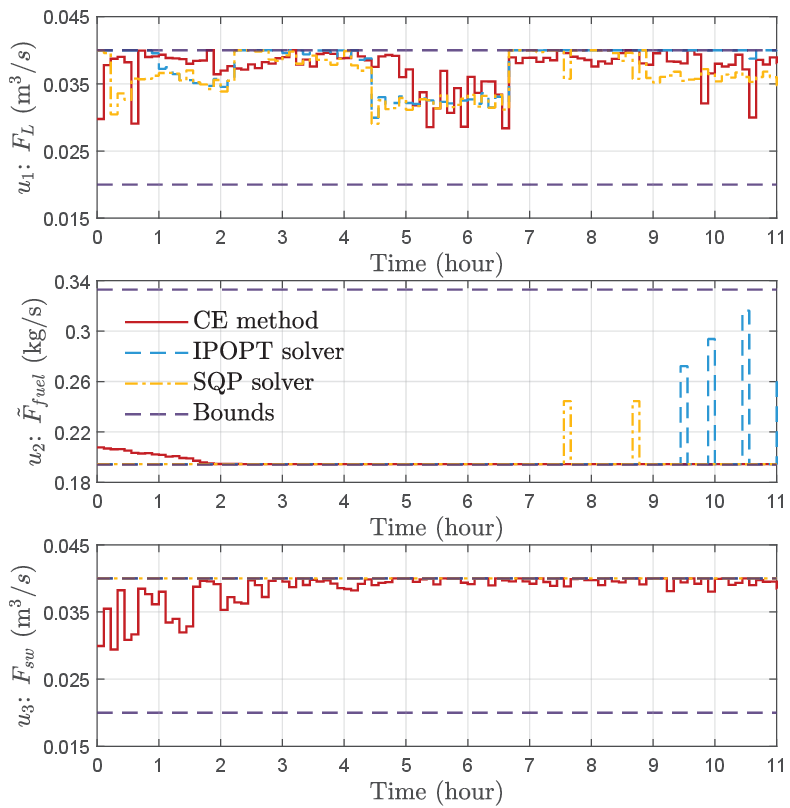}
    \caption{Closed-loop trajectories of control input obtained from hybrid model-based EMPC solved by the CE method~\cite{wen2018constrained,liu2020constrained}, the IPOPT solver~\cite{wachter2006implementation}, and the SQP solver~\cite{matlab}.}\label{fig:control:u}
\end{figure}

\begin{table}[!t]
    \renewcommand\arraystretch{1.2}
    \caption{Average one-step computation time across different control horizons using the CE method~\cite{wen2018constrained,liu2020constrained}, the IPOPT solver~\cite{wachter2006implementation}, and the SQP solver~\cite{matlab}.}\label{table:time}
    \centering\footnotesize%\small
      \begin{tabular}{c c c c c}
        \toprule
             &$N_c=5$ &$N_c=10$ &$N_c=15$ & $N_c=20$ \\ \midrule
        CE~\cite{wen2018constrained,liu2020constrained} (s/step)  &56  &349   &614  &1033  \\
        SQP~\cite{wachter2006implementation} (s/step)   &75  &435  &1108   &1857\\
        IPOPT~\cite{matlab} (s/step)   &483  &975  &2118  &4056 \\  \bottomrule
      \end{tabular}
  \end{table}

\textcolor{black}{First, we compare the control performance obtained based on the three optimization solvers under the same setting as mentioned at the beginning of this section, with the control horizon $N_c=5$. Figure~\ref{fig:control:ey} presents the closed-loop trajectories of economic cost rate, carbon capture rate, and the controlled outputs generated by the three optimization solvers. Figure~\ref{fig:control:u} shows the corresponding control input trajectories obtained using these three optimization solvers. The closed-loop trajectories generated by the three optimization solvers exhibit similar trends, with minor deviations, particularly in the economic cost rate, which is the primary objective of the EMPC optimization problem. The average economic cost rates based on results given by the CE method~\cite{wen2018constrained,liu2020constrained}, the IPOPT solver~\cite{wachter2006implementation}, and the SQP solver~\cite{matlab} are \$0.262/s, \$0.265/s, and \$0.263/s, respectively. %The difference in the economic cost rates given by the CE method between the IPOPT solver and the SQP solver are 1.13\% and 0.38\%, respectively.
When the IPOPT solver~\cite{wachter2006implementation} and the SQP solver~\cite{matlab} are used for solving the optimization for the proposed hybrid model-based EMPC method, several instances of significantly high economic cost rates over time are observed; this may stem from failures or the infeasiblity of th corresponding method in solving the optimization problem at the specific time instants.}

\textcolor{black}{Next, we compare the average computation time as we use the three optimization solvers considering different control horizons $N_c$, including 5, 10, 15, and 20. Table~\ref{table:time} presents the average computation times for the one-step execution of the three solvers. Specifically, the CE method~\cite{wen2018constrained,liu2020constrained} shows a reduction of 88.41\%, 64.21\%, 71.01\%, and 74.53\% in computation times for control horizons of 5, 10, 15, and 20, respectively, compared to the IPOPT solver~\cite{wachter2006implementation}. In comparison with the SQP solver~\cite{matlab}, the CE method~\cite{wen2018constrained,liu2020constrained} shows a reduction of 25.33\%, 19.77\%, 44.58\%, and 44.37\% in computation times for control horizons of 5, 10, 15, and 20, respectively.}

  % Moreover, we compare the computation time using the cross-entropy method with the interior point optimizer (IPOPT) solver~\cite{wachter2006implementation} for the proposed EMPC design. The average one-step computation times for implementing {the proposed method} using the CE method and IPOPT are 224 and 740 seconds, respectively. The computation times of the CE method are 69.73\% faster than those of IPOPT when solving the proposed EMPC based on the hybrid model.

\section{Conclusion}

In this paper, we introduced a novel approach to optimize the efficiency and economic operational performance of shipboard post-combustion carbon capture processes, which integrates both hybrid modeling and economic control strategies. We considered an integrated plant that encompasses the ship engine system and the shipboard post-combustion carbon capture plant. To robustly characterize the dynamic behaviors of the entire process, we developed a hybrid model that integrates imperfect first-principles with two neural networks trained using process data. Leveraging this hybrid model, an economic model predictive control (EMPC) scheme was formulated. We employed the cross-entropy method to tackle the online computation challenges posed by the non-convex optimization inherent in EMPC. Through extensive simulations and analysis, we evaluated the efficacy of our proposed approach. Our hybrid model outperforms solely data-based machine learning models -- the results demonstrated superior modeling performance, enhanced data efficiency, and robustness. Furthermore, it exhibited good generalization capabilities, in the sense that it accurately predicted process dynamics for operational conditions that were not considered during the model training phase. The control results demonstrated the effectiveness of the proposed EMPC scheme in minimizing the overall economic cost for real-time process operation, providing an 8.07\% cost reduction as compared to conventional optimal set-point tracking control based on the same hybrid model. \textcolor{black}{Additionally, the proposed hybrid model-based EMPC achieves a 4.20\% reduction in economic cost and a 9.10\% improvement in carbon capture rate compared to the imperfect first-principles model-based EMPC. By utilizing the cross-entropy method~\cite{wen2018constrained,liu2020constrained}, the computational complexity associated with nonlinear optimization within the EMPC framework is significantly reduced as compared to two nonlinear optimization solvers: the IPOPT solver~\cite{wachter2006implementation} and the SQP solver~\cite{matlab}.}

\section*{Acknowledgment}
This research is partially supported by Ministry of Education, Singapore, under its Academic Research Fund Tier 1 (RS15/21), and is partly supported by the National Research Foundation of Singapore and Singapore Maritime Institute (SMI) under the Maritime Transformation Program White Space Funding Support (SMI-2022-MTP-03) jointly with the Maritime Energy and Sustainable Development Centre of Excellence, Nanyang Technological University.

\section*{Appendix}

The process variables of the shipboard post-combustion carbon capture plant are listed in Table~\ref{paper1:table:process variables}. The symbols used in the hybrid modeling and economic model predictive control scheme are presented in Table~\ref{paper1:table:control variables}.

\renewcommand{\thetable}{A.1}

% \begin{table}[!t]
    \footnotesize
    \renewcommand\arraystretch{1.1}
    \captionof{table}{Process variables of the shipboard post-combustion carbon capture plant.}\label{paper1:table:process variables} \vspace{0pt} 
    \centering
    \begin{tabularx}{0.49\textwidth}{m{0.08\textwidth} m{0.365\textwidth}}
    
        \toprule
        \textbf{Variables} & \textbf{Meaning} \\
        \midrule
    % \begin{longtable}{m{5.8cm} m{9.9cm}}
    % \caption{Process variables of the shipboard post-combustion carbon capture plant.}\label{paper1:table:process variables} \vspace{-5pt} \\ \hline
    % \textbf{Variables} & \textbf{Meaning} \\
    % \hline
    % \endfirsthead
    % \multicolumn{2}{c}%
    % {\tablename\ \thetable: \textit{\rm{Process variables of the shipboard post-combustion carbon capture plant.}}} \\
    % \hline
    % \textbf{Variables} & \textbf{Meaning} \\
    % \hline
    % \endhead
    % \hline \multicolumn{2}{r}{\textit{Continued on next page}} \\
    % \endfoot
    % \hline
    % \endlastfoot
    % \small
          $a^I$ & Interfacial area in m$^2$/m$^3$\\
          $C_{L,i}/C_{G,i}$ &Molar concentration in liquid/gas phase of component $i$ in kmol/m$^3$, $i = \text{CO}_2, \text{MEA}, \text{H}_2\text{O}, \text{N}_2$\\
          $C_{p,i}$ &Heat capacity of component $i$ in kJ/(kmol$\cdot$K), $ i = \text{CO}_2, \text{MEA}, \text{H}_2\text{O}, \text{N}_2$ \\
          $\tilde{C}_{p,sw}/\tilde{C}_{p,sol}$ &Seawater/solvent heat capacity in kJ/(kg$\cdot$K)  \\
          $\tilde{C}_{p,flue}$ &Heat capacity of the flue gas in kJ/(kg$\cdot$K)  \\
          $\hat{C}_{p,reb}$ &Liquid average heat capacity in reboiler in kJ/(kmol$\cdot$K) \\
          $D_c$ &Internal  diameter of the column in m \\
          $F_L/F_G$ &Liquid/gas phase volumetric flow rate in m$^3$/s\\
          $F_{G,reb}$ &Gas phase volumetric flow rate exiting the reboiler in m$^3$/s \\
          $F_{sw}$ &Seawater volumetric flow rate in m$^3$/s\\
          $\hat{F}_{in}$ &Inlet stream flow rate of reboiler in kmol/s \\
          $\hat{F}_{L}/\hat{F}_V$ &Liquid/vapor flow rate of reboiler in kmol/s \\
          $\tilde{F}_{fuel}$ &Mass flow rate of fuel in kg/s \\
          $\tilde{F}_{\text{CO}_2}$ &CO$_2$ mass flow rate in treated gas in kg/s \\
          $\tilde{F}_{flue,\text{CO}_2}$ &Mass flow rate of CO$_2$ in flue gas in kg/s \\
          $H_{L,in}/H_{L,out}$ &Liquid enthalpy at the inlet/outlet in kJ/kmol \\
          $H_{V,out}$ &Vapor enthalpy leaving the reboiler in kJ/kmol \\
          $\hat{H}_{steam}/\hat{H}_{water}$ &Specific enthalpy of saturated steam/water at pressure 6 barG in kJ/kg\\
          $l$ &Length of the column in m \\
          $m_{i,in}/m_{i,out}$ &Inlet/outlet liquid  mole fraction of component $i$ of reboiler, $i = \text{CO}_2, \text{MEA}, \text{H}_2\text{O}$ \\
          $M_i$ &Mass holdup of component $i$ in reboiler in kmol, $i = \text{CO}_2, \text{MEA}, \text{H}_2\text{O}$ \\
          $N_i$ &Mass transfer rate of component $i$ in kmol/(m$^2\cdot$s), $i = \text{CO}_2, \text{MEA}, \text{H}_2\text{O}, \text{N}_2$ \\
          $q_{i, out}$ &Outlet vapor mole fraction of component $i$ of reboiler \\
          $q_{reb}$ &Vapor fraction of reboiler, $i = \text{CO}_2, \text{MEA}, \text{H}_2\text{O}$ \\
          $q_{flue, \text{CO}_2}$ &Gas mole fraction of CO$_2$ in flue gas \\
          $q_{fuel,\text{C}}$  & Mole fraction of carbon in the fuel \\
          $Q_L/Q_G$ &Liquid/gas phase interfacial heat transfer rate in kW/m$^2$ \\
          $Q_E$ &Total engine power in kW \\
          $Q_{reb}$ &Reboiler heat duty in kW \\
          $Q_{turbine}$ &Power generated by diesel gas turbine in kW \\
          $Q_{rec}$ &Power generated by WHR in kW \\
          $r_{\text{CO}_2}$ &Molar mass of CO$_2$ in kg/kmol \\
          $r_{\text{C}}$ &Molar mass of carbon in kg/kmol \\
          $T_L/T_G$ &Liquid/gas phase temperature in K \\
          $T_{sol, in}/T_{sol, out}$ &Inlet/outlet solvent temperature of the seawater heat exchanger in K \\
          $T_{sw, in}/T_{sw, out}$ &Inlet/outlet seawater temperature of the seawater heat exchanger in K \\
          $T_{rec,in}/T_{rec,out}$ &Inlet/outlet temperature of the WHR system in K\\
          $T_{reb}$ &Reboiler temperature in K \\
          $U$ &Overall heat transfer coefficient of lean-rich solvent heat exchanger in kW/K \\
          $V_{reb}$ &Holdup volumne in reboiler in m$^3$ \\
          $W_{SFOC}$ &Specific fuel oil consumption in kg/kWh \\
    
          $\rho_{reb}$ &Liquid density in the reboiler in kmol/m$^3$ \\
          $\hat{\rho}_{sw}/\hat{\rho}_{sol}$ &Seawater/solvent density in kg/m$^3$ \\
          $\tilde{\rho}_{flue}$ & Flue gas density in kg/m$^3$ \\
          $\eta_{fuel}$ &Calorific value of fuel in kJ/kg \\
          $\varphi_E$ &Engine load ratio \\
          $\varphi_{\text{CO}_2}$ & Carbon capture rate \\
          \bottomrule
    % \end{longtable}
    \end{tabularx}
% \end{table}
% Change to to normal numbering
\renewcommand{\thetable}{\arabic{chapter}.\arabic{table}}

\begin{center}\vspace{30pt}
    % \renewcommand\arraystretch{1.13}
    % \LTcapwidth = \textwidth
    % \setstretch{1.1}
    % \begin{longtable}{|m{5.8cm}|m{9.8cm}|}
    % Change to Appendix A.1
    \renewcommand{\thetable}{A.2}

    % \begin{table}[!t]
        \footnotesize
        \renewcommand\arraystretch{1.1}
        \captionof{table}{Symbols of hybrid modeling and economic model predictive control scheme.}\label{paper1:table:control variables}\vspace{-3pt} 
        \centering
        \begin{tabularx}{0.49\textwidth}{m{0.05\textwidth} m{0.395\textwidth}}
        
            \toprule
            \textbf{Variables} & \textbf{Meaning} \\
            \midrule
    % \begin{longtable}{m{3cm} m{11cm}}
    % \caption{Symbols of hybrid modeling and economic model predictive control scheme.} \vspace{-5pt} \\ \hline
    % \textbf{Variables} & \textbf{Meaning} \\
    % \hline
    % \endfirsthead
    % \multicolumn{2}{c}%
    % {\tablename\ \thetable: \textit{\rm{Symbols of hybrid modeling and economic model predictive control scheme.}}} \\
    % \hline
    % \textbf{Variables} & \textbf{Meaning} \\
    % \hline
    % \endhead
    % \hline \multicolumn{2}{r}{\textit{Continued on next page}} \\
    % \endfoot
    % \hline
    % \endlastfoot
    % \small
      $x$  &Differential states including 103 variables\\
      $z$  &Algebraic states including 7 variables \\
      $u$  &Control inputs including 3 variables\\
      $p$  &Known disturbance including 1 variable\\
      $y$  &Controlled outputs including 2 variables \\
      $\tilde{x}^{fp}$ &Differential states predicted by imperfect first-principles model \\
      % $\tilde{z}^{fp}$ &Algebraic state predicted by first-principles model \\
      $\hat{z}^{nn}$ &Algebraic states predicted by algebraic state inference\\
      $\hat{x}^{nn}$ &Differential state error predicted by state dynamics compensation \\
      $\hat{x}$ &Differential states predicted by hybrid model \\
      $\hat{y}$ &Controlled outputs predicted by hybrid model \\
      $y_{limit}$ &Limit value for CO$_2$ emissions in kg/s \\
      \textcolor{black}{$\mathbf{X}$} &\textcolor{black}{Open-loop data of differential states generated by simulator} \\
      \textcolor{black}{$\mathbf{Z}$} &\textcolor{black}{Open-loop data of algebraic states generated by simulator} \\
      \textcolor{black}{$\mathbf{U}$} &\textcolor{black}{Open-loop data of control inputs generated for modeling} \\
      \textcolor{black}{$\mathbf{P}$} &\textcolor{black}{Open-loop data of the known disturbance generated for modeling} \\
      \textcolor{black}{$\mathbf{\tilde{\mathbf{X}}^+}$} &\textcolor{black}{One-step-ahead prediction of differential states by imperfect first-principles model} \\
      \textcolor{black}{$\mathbf{X}^+_e$} &\textcolor{black}{One-step mismatch between the differential states of the simulator and the imperfect first-principles model} \\
      $\nu_{min}$ &Minimum variance bound \\
      % $\theta_{G_{nn}}^*$ &Well-trained parameters of algebraic state inference \\
      % $\theta_{F_{nn}}^*$ &Well-trained parameters of state dynamics compensation \\
      %$\mathcal{L}_z(\theta_{G_{nn}})$ &Loss of algebraic state inference \\
      %$\mathcal{L}_x(\theta_{F_{nn}})$ &Loss of state dynamics compensation \\
      %$\mathcal{D}_z$ &Data set of algebraic state inference \\
      %$\mathcal{D}_x$ &Data set of state dynamics compensation \\
      $\mathcal{J}$ &Accumulated cost\\
      \textcolor{black}{$Q_{mpc}$} & \textcolor{black}{Weighting matrices for the outputs of MPC}\\
      \textcolor{black}{$R_{mpc}$} & \textcolor{black}{Weighting matrices for the control inputs of MPC} \\
      $N_{epochs,z}$ &Training epoch number of algebraic state inference \\
      $N_{epochs,x}$ &Training epoch number of state dynamics compensation \\
      $N_{dim}$ &Number of the state dimensions \\
      $N_{data}$ &Number of data samples \\
      $N_c$ &Control horizon of EMPC \\
      $N_{CE}$ & Iteration number of CE method~\cite{wen2018constrained,liu2020constrained} \\
      $N_{sample}$ &Number of generated samples per iteration of CE method~\cite{wen2018constrained,liu2020constrained} \\
      $N_K$ & Size of elite samples \\
      $\alpha$ &Carbon tax in $\$$/kg \\
      $\beta$ &Fuel price in $\$$/kg \\
      $\theta_{G_{nn}}$ &Trainable parameters of algebraic state inference \\
      $\theta_{F_{nn}}$ &Trainable parameters of state dynamics compensation \\
      $\lambda$ & Update rate of CE method~\cite{wen2018constrained,liu2020constrained} \\
       \bottomrule
% \end{longtable}
\end{tabularx}
% \end{table}
% Change to to normal numbering
\renewcommand{\thetable}{\arabic{chapter}.\arabic{table}}
\end{center}

% % \bibliographystyle{Unsrt}
% \bibliographystyle{unsrt}
% % \bibliographystyle{ieeetr}
% % \bibliography{Refv4}
% \bibliography{Refv4}
% % \printbibliography

% \newpage
% \clearpage
% \FloatBarrier

% \raggedleft
\normalsize
\bibliographystyle{plain}

\end{document}